\documentclass[a4paper,fleqn]{cas-dc}
\catcode `\@11\relax 
\catcode `\_11\relax
\catcode `\:11\relax
\def\__first_footerline:{\group_begin: \small \sffamily \ifnum \theblind >0\relax \else 
  \__short_authors: : \fi {\rmfamily \itshape Preprint submitted}\group_end:
}

\catcode `\@12\relax 
\catcode `\_8\relax
\catcode `\:12\relax

\usepackage[authoryear]{natbib}

\usepackage{amsmath, amssymb, amsfonts, graphicx,color,mathtools,amsthm,xcolor}
\usepackage{graphicx}
\usepackage{caption, subcaption}
\usepackage{algorithm, algorithmic}
\usepackage{url}
\usepackage{tikz}
\usepackage{placeins}
\usepackage{booktabs}
\usepackage{tabularx}
\usepackage{mathrsfs}
\usepackage{microtype}
\usepackage{hyperref}
\usepackage{float}
\usepackage{csvsimple}
\usepackage{array}
\usepackage{natbib}
\usepackage{xurl}

\newcolumntype{C}[1]{>{\centering\let\newline\\\arraybackslash\hspace{0pt}}m{#1}}

\def\tsc#1{\csdef{#1}{\textsc{\lowercase{#1}}\xspace}}
\tsc{ZY}
\tsc{GPTC}
\tsc{LQ}

\begin{document}
\let\WriteBookmarks\relax
\def\floatpagepagefraction{1}
\def\textpagefraction{.001}
\shorttitle{Geometric quantification and shape analysis for axillary lymph node metastasis}
\shortauthors{Z. Yi et~al.}

\title [mode = title]{An integrated geometric quantification and shape analysis framework for axillary lymph node metastasis in breast cancer patients}                      
\tnotemark[1]

\tnotetext[1]{This work is supported by the Croucher Tak Wah Mak Innovation Award (to GPTC) and the Scientific Research Program of FuRong Laboratory (No. 2025PT5047).}


\author[1]{Zixi Yi}

\credit{Conceptualization, Data curation, Formal analysis, Investigation, Methodology, Software, Validation, Visualization, Writing – original draft}

\affiliation[1]{organization={School of Mathematics and Statistics, Central South University},
                city={Changsha, Hunan},
                country={China}}

\author[2,3]{Limeng Qu}
\credit{Data curation, Methodology, Writing – review \& editing}

\affiliation[2]{organization={Department of General Surgery, The Second Xiangya Hospital, Central South University},
                city={Changsha, Hunan},
                country={China}}

\affiliation[3]{organization={Clinical Research Center for Breast Disease in Hunan Province},
                city={Changsha, Hunan},
                country={China}}

\author[4]{Gary P. T. Choi}[orcid=0000-0001-5407-9111]
\ead{ptchoi@cuhk.edu.hk}
\cormark[1]
\credit{Conceptualization, Funding acquisition, Investigation, Methodology, Project administration, Supervision, Writing – review \& editing}

\affiliation[4]{organization={Department of Mathematics, The Chinese University of Hong Kong},
                city={Hong Kong SAR},
                country={China}}

\cortext[cor1]{Corresponding author}

\begin{abstract}
Quantitative characterization of lymph node morphology is important for assessing axillary lymph node metastasis in breast cancer. However, surfaces reconstructed from computed tomography (CT) segmentation may contain geometric and topological defects that compromise subsequent analysis, while conventional shape descriptors predominantly characterize global morphology. To address these issues, we developed an integrated framework combining topology-aware surface processing with multi-resolution spherical harmonic (SH) analysis of CT-derived axillary lymph nodes. The processing pipeline produced topology-valid genus-0 surfaces with improved mesh quality, which were then represented at multiple SH degrees and characterized using 20 predefined geometric feature families. Geometric fidelity increased with SH degree, whereas predictive performance peaked at intermediate resolutions, indicating that maximal reconstruction fidelity did not coincide with maximal discriminative utility. Preferred SH degree also differed across feature families. A family-specific mixed-resolution model achieved an AUC of 0.918 (95\% CI, 0.904--0.934), compared with 0.884 (95\% CI, 0.868--0.900) for the conventional PyRadiomics Shape14 baseline, corresponding to an improvement of 0.0344 (95\% paired bootstrap CI, 0.0212--0.0468). Controlled perturbation experiments showed that higher SH degrees transmitted more fine-scale geometric variation and yielded lower stability of curvature-based predictions. Representative geometric descriptors provided interpretable characterization of metastasis-associated surface morphology. Independent validation further supported the transportability of the framework: label-free replication in a multicenter lymph node cohort reproduced the family-specific resolution effects, while a labeled LIDC-IDRI lung-nodule experiment reproduced the resolution-dependent relationship between SH degree and predictive performance. Altogether, the proposed framework provides a topology-valid basis for multi-scale quantitative characterization of lymph node morphology and metastasis-associated imaging phenotypes.

\end{abstract}




\maketitle

\section{Introduction}
Axillary lymph node metastasis is a key component of regional staging in breast cancer and has important implications for locoregional treatment planning and prognosis~\citep{marino2020lymph}. Quantitative imaging and radiomics provide a means of characterizing disease-related phenotypes beyond qualitative visual assessment \citep{lambin2017radiomics}, and CT-derived lymph node features have demonstrated predictive value for metastatic status in both single-center and multicenter studies \citep{yang2021ct,qu2024unsupervised,qu2025evaluating}. In clinical imaging, metastatic lymph nodes may exhibit cortical thickening, focal cortical bulging, loss of the fatty hilum, and a transition from an oval or reniform shape toward a round, asymmetric, or irregular contour. These morphological changes may reflect partial or diffuse replacement of the nodal architecture by metastatic tumor, with capsular invasion and extranodal extension occurring in more advanced nodal involvement~\citep{dialani2018indications}.
However, lymph node morphology is commonly represented by image-derived radiomic variables or a limited set of global shape descriptors describing size, compactness, principal-axis geometry, and spatial extent. Such measurements may obscure localized or asymmetric changes and provide limited characterization of local surface geometry, symmetry, and spatially heterogeneous three-dimensional morphology. Metastatic involvement may alter nodal morphology in a spatially heterogeneous manner, producing asymmetric enlargement, focal contour bulging, lobulation, or contour irregularity that may not be fully reflected by global measurements of size, elongation, or compactness.

Explicit surface analysis introduces an additional computational requirement: meshes reconstructed from CT-derived lymph node segmentations must provide a numerically and topologically valid domain for subsequent geometric operators. Limited spatial resolution, anisotropic voxel sampling, segmentation boundaries, and surface reconstruction can introduce staircase artifacts, local geometric irregularities, disconnected structures, and topological inconsistencies \citep{Skrinjar2009,Moench2011,Ito2019,Shirshin2021}. Existing smoothing and topology-repair techniques address individual failure modes \citep{taubin1995signal,desbrun1999implicit,chen2025topomesh}, but smoothing alone does not guarantee a valid spherical domain, and genus alone does not ensure connectedness, closedness, manifoldness, or consistent orientation. SH-based reconstruction has also been explored for correcting topological defects
in medically derived surface meshes, particularly for cortical surfaces
\citep{yotter2009topological}. Reliable surface-based analysis therefore requires coordinated control of geometric artifacts, topology, mesh quality, and geometric preservation.

Once a valid genus-$0$ surface has been obtained, spherical parameterization provides a natural common domain for representing and comparing three-dimensional geometry \citep{choi2015flash,lyu2024spherical}. Spherical harmonics (SH) are particularly well suited to this setting because they provide an ordered spectral representation in which surface geometry can be reconstructed at progressively increasing spatial resolutions \citep{brechbuehler1995parametrization,styner2006spharm,khairy2008spherical}. Recent work has further investigated improved SH sampling and reconstruction strategies for general three-dimensional shape representation \citep{li2024fsh3d}. Low harmonic degrees primarily encode coarse, large-scale morphology, whereas higher degrees progressively recover finer geometric structure. SH therefore provide two complementary advantages for quantitative surface analysis: a compact representation of genus-$0$ geometry and an explicit mechanism for controlling the spatial scale at which morphology is represented.

SH-derived representations have previously been used for medical shape classification, including three-dimensional lung-nodule malignancy analysis and myocardial-infarction screening \citep{elbaz2011lung,valizadeh2021parametric}. However, these studies primarily used SH-derived representations or coefficients as predictive shape information rather than systematically examining how the finite reconstruction bandwidth influences downstream geometric measurements. This scale dependence raises a central methodological question: the SH degree required for faithful geometric reconstruction need not be the degree that most effectively represents morphology associated with metastatic status. Increasing the maximum harmonic degree expands representational capacity and improves the recovery of fine surface detail, but the additional high-frequency information may contain both disease-related morphology and variation that is only weakly associated with the clinical endpoint. Conversely, truncation at a lower or intermediate degree may suppress fine-scale variation while preserving larger-scale morphological organization. Geometric fidelity and discriminative utility are therefore distinct properties of an SH representation, and optimizing one does not necessarily optimize the other. Resolution may also influence robustness, because higher SH degrees can retain finer geometric perturbations arising from segmentation uncertainty, reconstruction artifacts, or other small surface variations, whereas spectral truncation may attenuate such perturbations. SH degree can therefore be viewed not only as a reconstruction parameter but also as a candidate geometric scale governing the balance among geometric fidelity, discriminative information, and sensitivity to perturbation.

The appropriate geometric scale may further depend on the quantity being measured. Conventional three-dimensional shape features, such as those defined in the PyRadiomics framework, primarily summarize global size, principal-axis geometry, compactness, and surface-to-volume relationships \citep{vangriethuysen2017pyradiomics}. Differential-geometric descriptors characterize local surface bending and shape type \citep{meyer2003discrete,koenderink1992surface}; Laplace--Beltrami spectra and diffusion-based signatures characterize intrinsic geometry across multiple scales \citep{reuter2006shapedna,sun2009hks,aubry2011wks}; geodesic measures describe intrinsic spatial extent \citep{crane2013geodesics}; and shape-distribution and symmetry descriptors capture complementary aspects of three-dimensional spatial organization \citep{osada2001,kazhdan2004}. Because these descriptors depend on different geometric properties and spatial neighborhoods, there is little reason to expect a single SH degree to represent all feature families equally effectively. A uniformly low resolution may suppress information required by locally sensitive descriptors, whereas a uniformly high resolution may retain fine-scale variation that contributes little to more global measurements or to discrimination. This motivates a feature-dependent treatment of reconstruction scale rather than the assumption of a universally optimal SH degree.

In this study, we develop an integrated framework for topology-aware processing and multi-resolution geometric analysis of CT-derived axillary lymph node surfaces. We first establish a processing workflow that combines context-aware smoothing, adaptive topology correction, geometry-constrained smoothing, and local artifact reconstruction, and evaluate the resulting surfaces in terms of topological validity and computational quality. We then examine SH representations across multiple harmonic degrees to determine how reconstruction resolution affects geometric fidelity, spectral discrimination, and predictive performance. Geometric descriptors are organized into predefined feature families to evaluate feature-dependent resolution preferences and construct a family-specific mixed-resolution representation. Resolution-dependent robustness is further assessed through geometric perturbation transmission, harmonic-domain perturbation energy, and downstream predictive stability. Finally, independent multicenter lymph node and cross-organ LIDC-IDRI experiments are used to assess transportability of the observed geometric and resolution-dependent effects.

\section{Data and geometric processing of lymph node surfaces}
\label{sec:data_geometric_processing}

\subsection{Study cohort and input data}

The CT-derived lymph node label maps analyzed in this study were obtained from previously reported datasets~\citep{qu2024unsupervised,qu2025evaluating}. Specifically, the published Semi-ALNP cohort included 214 women with unilateral invasive breast cancer who underwent preoperative high-resolution thin-section contrast-enhanced CT and axillary lymph node dissection (ALND). The main eligibility criteria were the absence of distant metastasis and no neoadjuvant therapy before imaging. Briefly, individual axillary lymph nodes were delineated on consecutive thin-slice contrast-enhanced CT images. CT was performed within 1 month before surgery using a high-resolution thin-section contrast-enhanced protocol, with a 1-mm reconstructed slice thickness and 0-mm spacing. Histopathological examination of the ALND specimens served as the reference standard. The labels were defined using two patient groups: 1,769 ALNs from 214 patients, including 1,172 ALNs from 161 patients with no nodal metastasis and 597 ALNs from 53 patients in whom all dissected ALNs were metastatic. In the earlier dataset, three-dimensional lymph node localization was assisted by VitaWorks, followed by segmentation in 3D Slicer. In the subsequent Semi-ALNP dataset, ALNs were segmented in ITK-SNAP v3.8.0 using a region-growing procedure by two independent operators who were blinded to clinical and pathological information. The segmentations were subsequently reviewed and finalized by a senior radiologist. These procedures yielded complete three-dimensional lymph node label volumes.

An independent multicenter external-validation cohort was retrospectively collected from four hospitals between January 2023 and July 2024~\citep{qu2025evaluating}. The external dataset comprised CT-derived three-dimensional axillary lymph node label maps from patients with breast cancer. Individual CT-visible lymph nodes in this cohort were not linked to node-level histopathological metastasis labels; therefore, metastatic status was unavailable for the external surfaces included in the present analysis. An additional cross-organ validation experiment was performed using segmented lung nodules from the publicly available LIDC-IDRI dataset~\citep{armato2011lung}. After application of the predefined surface-availability and analysis criteria, 59 lung nodules with corresponding classification labels were retained. This cohort was analyzed independently from the axillary lymph node datasets to examine whether the relationship between SH reconstruction degree and predictive performance was preserved in a distinct anatomical structure and downstream classification task. The LIDC-IDRI data were not used for the development of the lymph node prediction models, the selection of lymph node feature families, or the determination of the SH degrees examined in the primary analysis.

For all cohorts, the resulting NIfTI label maps were converted into triangular surface meshes using a smoothing factor of $0.2$. This setting provided mild suppression of voxel-scale irregularities while limiting excessive surface deformation and the introduction of additional geometric artifacts.

\subsection{Overview of the geometric processing workflow}
\label{sec:workflow_overview}

\begin{figure*}
    \centering
    \includegraphics[width=\linewidth]{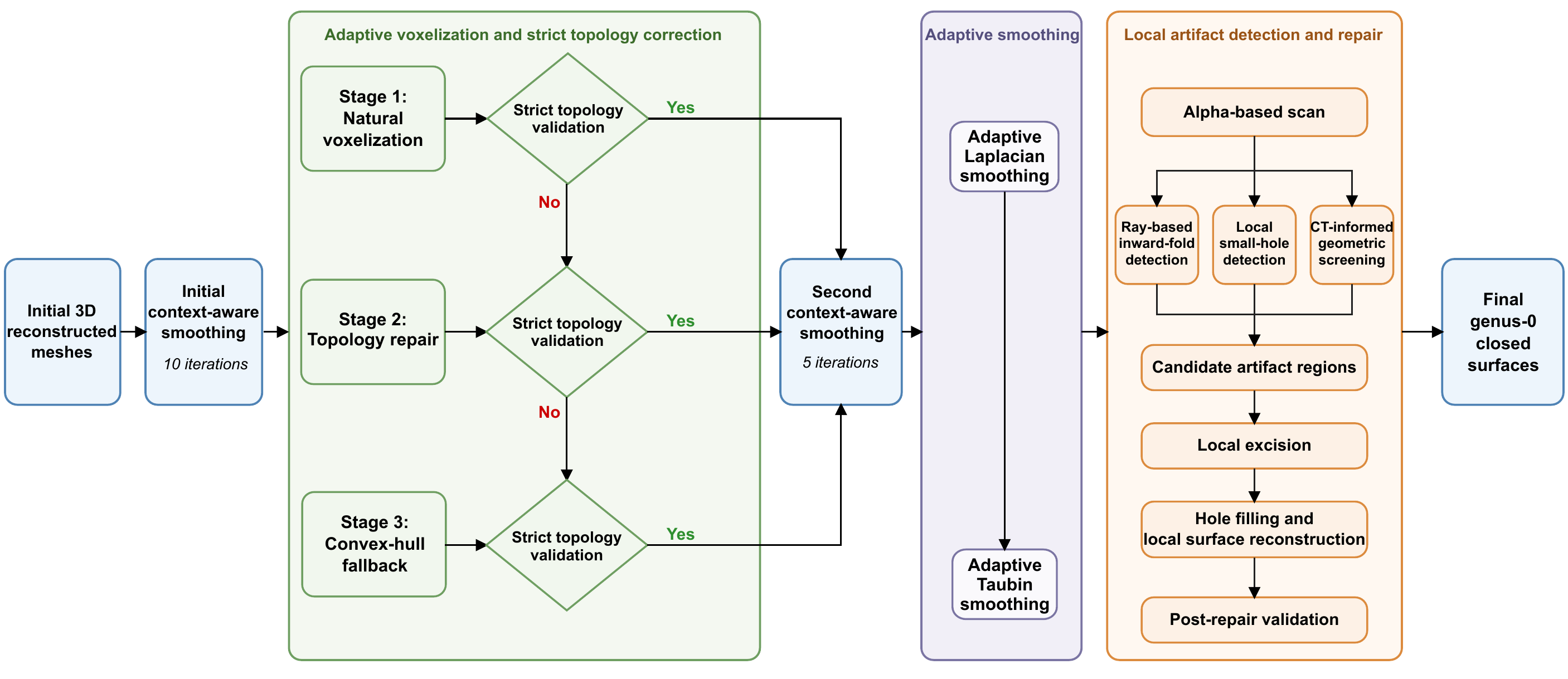}
    \caption{\textbf{Overview of the proposed geometric processing workflow.}
    The pipeline comprises initial context-aware smoothing, adaptive voxelization with strict topology correction, geometry-constrained adaptive smoothing, and local artifact detection and repair. The post-repair validation criteria ensure that the output surfaces are connected, closed, consistently oriented genus-$0$ surfaces for the subsequent spherical analysis.}
    \label{fig:processing_workflow}
\end{figure*}

The reconstructed lymph node surfaces were processed using the sequential geometric workflow illustrated in Fig.~\ref{fig:processing_workflow}. The workflow was designed to suppress segmentation- and reconstruction-related artifacts, enforce the topological requirements for spherical parameterization, and improve local mesh quality while limiting unnecessary modification of the underlying surface geometry.

The initial meshes underwent context-aware smoothing to attenuate staircase artifacts, followed by adaptive voxelization with strict topology validation. At each candidate voxel-grid resolution, natural voxelization was evaluated before explicit topology repair. A naturally voxelized surface was accepted whenever it satisfied all predefined validity criteria. Explicit topology repair was invoked only when no naturally valid candidate was available, and a convex-hull-based reconstruction was retained as a final fallback when necessary. A second context-aware smoothing stage was then applied to attenuate staircase-like irregularities introduced or retained during voxelization.

Residual geometric irregularities were further reduced using geometry-constrained adaptive Laplacian and Taubin smoothing. Local surface artifacts were subsequently identified using an initial $\alpha$-shape-based candidate scan followed by complementary filtering procedures targeting inward folds, small recessed defects, and CT-related geometric artifacts. The selected artifact regions were locally excised and reconstructed using boundary-aware triangulation and geometric fairing, after which the repaired surfaces underwent strict post-repair validation. Only connected, closed, consistently oriented genus-$0$ surfaces satisfying all predefined mesh-integrity criteria were retained for subsequent spherical parameterization and spherical harmonic analysis.

\subsection{Context-aware smoothing for staircase-artifact suppression}

To attenuate staircase artifacts while limiting deformation of unaffected surface regions, we applied the context-aware smoothing method of \cite{Moench2011}. The method identifies staircase-prone vertices from local variations in incident-face orientation and propagates a distance-dependent smoothing weight over the connected mesh neighborhood. Smoothing is therefore concentrated around detected artifacts rather than applied uniformly to the entire surface.

Using the displacement convention below, the local displacement at vertex $\mathbf{p}_i$ was
\begin{equation}
\mathbf{D}_i
=
\frac{1}{|\mathcal{N}_i|}
\sum_{j\in\mathcal{N}_i}
\left(
\mathbf{p}_i-\mathbf{p}_j
\right),
\label{eq:context_displacement}
\end{equation}
where $\mathcal{N}_i$ denotes the one-ring vertex neighborhood of $\mathbf{p}_i$. Each context-aware Taubin iteration consisted of two weighted updates:
\begin{align}
\mathbf{p}_i^{\left(k+\frac{1}{2}\right)}
&=
\mathbf{p}_i^{(k)}
-
w_i\lambda\mathbf{D}_i^{(k)},\\
\mathbf{p}_i^{(k+1)}
&=
\mathbf{p}_i^{\left(k+\frac{1}{2}\right)}
-
w_i\mu\mathbf{D}_i^{\left(k+\frac{1}{2}\right)},
\end{align}
with $\lambda=0.5$ and $\mu=-0.53$. The spatial weights $w_i$ were computed once at the beginning of each smoothing stage using the artifact-detection and distance-decay formulation of \cite{Moench2011}.

In our implementation, the staircase-detection threshold, maximum influence distance, and minimum nonzero smoothing weight were set to $\tau_s=0.55$, $d_{\max}=5$, and $w_{\min}=0.4$, respectively. Context-aware smoothing was applied for $10$ iterations before voxelization and reapplied for $5$ iterations afterward to suppress staircase-like irregularities introduced or retained during voxelization. Representative examples demonstrating the effect of context-aware smoothing are provided in Fig.~\ref{fig:surface_mesh_smoothing_comparison}.

\subsection{Adaptive voxelization and strict genus-\texorpdfstring{$0$}{0} topology correction}
\label{sec:adaptive_voxelization}

Spherical parameterization requires a connected, closed, consistently oriented genus-$0$ surface~\citep{choi2015flash}. However, meshes reconstructed from CT-derived lymph node segmentations may contain disconnected components, cavities, handles, or other topological defects. We therefore adapted the voxel-based mesh-repair framework used in previous geometry-image studies~\citep{sinha2016deep,pumarola20193dpeople} by introducing an adaptive, natural-first resolution search together with strict topology and mesh-integrity validation.

\subsubsection{Voxel-based reconstruction and strict validation}

For each candidate voxel-grid resolution $r$, the input triangular mesh was converted to a binary voxel representation. The largest $26$-connected occupied component was retained after local dilation and hole filling, and a three-dimensional $\alpha$-shape was reconstructed from the occupied voxel centers~\citep{edelsbrunner1994three}. The initial $\alpha$ parameter was $0.9$, and the resulting surface was transformed back to the coordinate system of the input mesh.

Candidate surfaces were evaluated using criteria stricter than genus alone. For a triangular mesh $\mathcal{M}$ with vertex, edge, and face sets $V$, $E$, and $F$, the Euler characteristic is
\begin{equation}
\chi(\mathcal{M})
=
|V|-|E|+|F|.
\label{eq:euler_characteristic}
\end{equation}
For a connected, closed, orientable surface,
\begin{equation}
g(\mathcal{M})
=
1-\frac{\chi(\mathcal{M})}{2}.
\label{eq:single_component_genus}
\end{equation}

Let $C$ denote the number of connected components and let $B$, $N$, $O$, $D$, $Z$, and $U$ denote the numbers of boundary edges, non-manifold edges, orientation conflicts, duplicate faces, degenerate faces, and unreferenced vertices, respectively. A candidate was considered strictly valid only when
\begin{equation}
\resizebox{.9\linewidth}{!}{$
\mathcal{Q}(\mathcal{M})
=
\begin{cases}
1,
&
C=1,\
g(\mathcal{M})=0,\
B=N=O=D=Z=U=0,\\
0,
& \text{otherwise}.
\end{cases}
$}
\label{eq:strict_topology_indicator}
\end{equation}
The genus was evaluated only after the mesh had been confirmed to be a closed, consistently oriented two-manifold. Surfaces failing these prerequisites were treated as topologically invalid.

\subsubsection{Adaptive natural-first search and topology repair}

The candidate voxel-grid resolutions were $\mathcal{R} = \{$150, 130, 110, 90, 70, 50$\}$, 
and were evaluated from the finest to the coarsest resolution.

At each resolution, natural voxelization was evaluated first, and the first candidate satisfying Eq.~(\ref{eq:strict_topology_indicator}) was retained. If no natural candidate was valid, topology repair was applied over the same resolution sequence; if this also failed, a convex-hull reconstruction was used as the final fallback, subject to the same strict validity criterion.

The strict topology and mesh-integrity validation was repeated immediately before export.

\begin{figure}
    \centering
    \includegraphics[width=\linewidth]{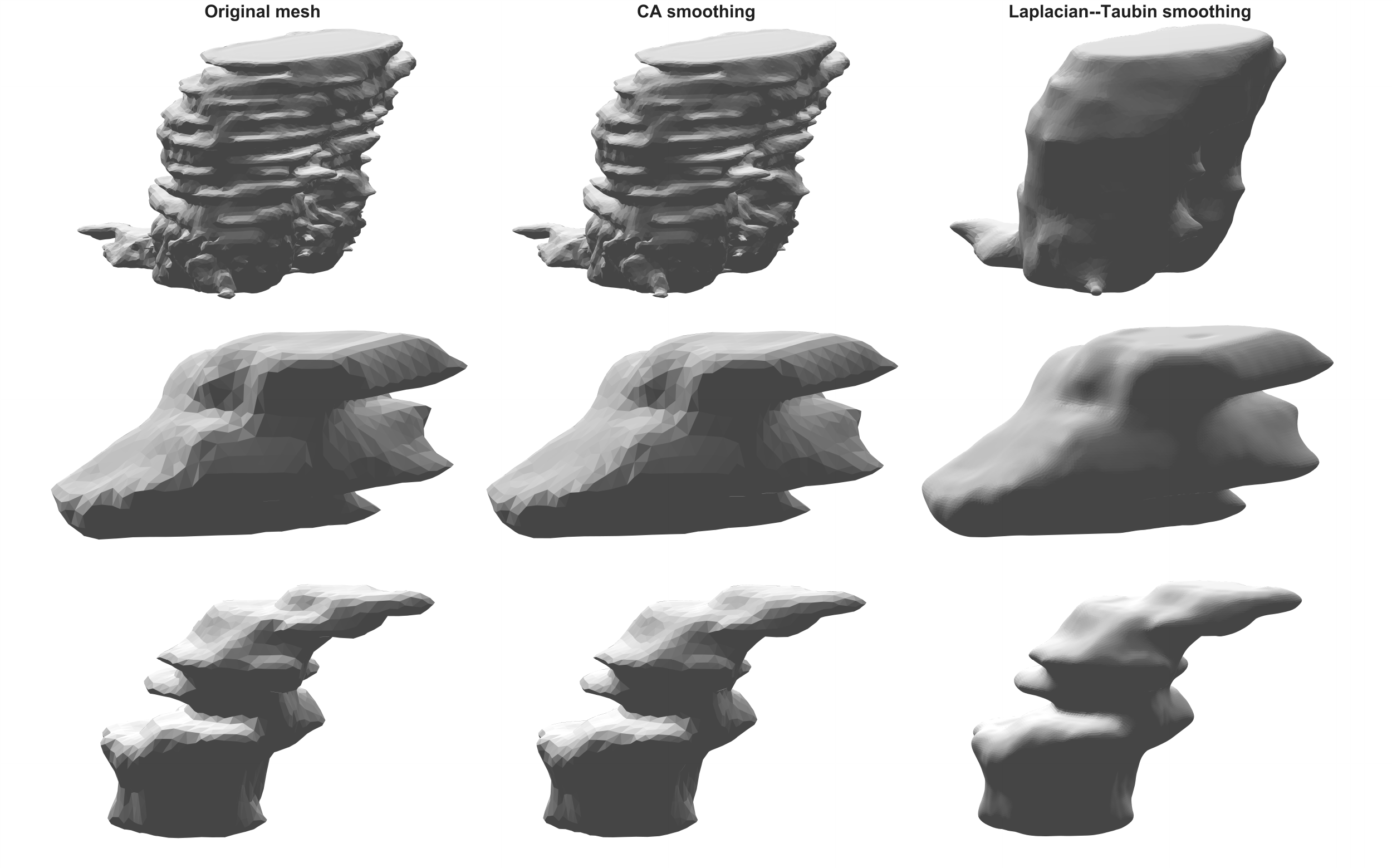}
    \caption{\textbf{Qualitative comparison of surface smoothing stages.}
    Representative lymph node surface meshes are shown before smoothing, after context-aware (CA) smoothing, and after adaptive Laplacian--Taubin smoothing.}
    \label{fig:surface_mesh_smoothing_comparison}
\end{figure}

\subsection{Geometry-constrained adaptive smoothing}
\label{sec:adaptive_smoothing}

Although the preceding processing steps corrected major topological defects and staircase-like artifacts, residual high-frequency geometric irregularities could remain. Standard Laplacian and Taubin smoothing can reduce such variation but may also alter the underlying geometry when applied excessively~\citep{taubin1995signal,desbrun1999implicit,vollmer1999improved}. Rather than applying a fixed smoothing duration to every surface, we developed a two-stage adaptive procedure in which the number of iterations was selected independently for each mesh according to improvement in surface smoothness subject to predefined geometric-fidelity constraints. As illustrated in Fig.~\ref{fig:surface_mesh_smoothing_comparison},
the subsequent geometry-constrained smoothing further attenuated residual
surface irregularities while preserving the overall morphology.

\subsubsection{Adaptive Laplacian stage}

The first stage used implicit cotangent-Laplacian smoothing~\citep{desbrun1999implicit,meyer2003discrete}. Using the negative-semidefinite Laplacian convention of the implementation, the update was
\begin{equation}
\left(
\mathbf{I}-\lambda_{\mathrm{L}}\mathbf{L}
\right)
\mathbf{X}^{(t+1)}
=
\mathbf{X}^{(t)},
\qquad
\lambda_{\mathrm{L}}=0.05.
\label{eq:implicit_laplacian_smoothing}
\end{equation}

Surface smoothness was quantified using the integrated squared Laplace--Beltrami energy
\begin{equation}
E_{\mathrm{L}}(\mathcal{M})
=
\sum_{i=1}^{n}
A_i
\left\|
\Delta_{\mathcal{M}}\mathbf{x}_i
\right\|^{2},
\label{eq:discrete_laplacian_energy}
\end{equation}
where $A_i$ is the lumped area associated with vertex $i$. 

Geometric fidelity was constrained by the relative enclosed-volume change and the maximum relative change among the three principal-axis extents, both measured with respect to the input surface.

\subsubsection{Adaptive Taubin stage}

The Laplacian-smoothed surface was subsequently refined using standard Taubin smoothing~\citep{taubin1995signal}, with $\lambda_{\mathrm{T}}=0.10$ and $\mu_{\mathrm{T}}=-0.11$. For adaptive stopping, local curvature variation was characterized using the absolute angle defect
$K_i=\left|2\pi-\sum_{f\in\mathcal{F}_i}\theta_{if}\right|$.
Following \cite{wang2012roughness}, the area-weighted global roughness was
\begin{equation}
R(\mathcal{M})
=
\frac{
\sum_{i=1}^{n}A_i\rho_i
}{
\sum_{i=1}^{n}A_i
},
\label{eq:global_curvature_roughness}
\end{equation}
where $\rho_i$ measures the difference between $K_i$ and its cotangent-weighted local neighborhood average.

Spatial deviation was additionally constrained using a symmetric, area-weighted $95$th-percentile Hausdorff distance following the mesh-based framework of MeshMetrics~\citep{podobnik2025meshmetrics}. The distance was normalized by the effective voxel scale,
\begin{equation}
h
=
\frac{\max\{\ell_x,\ell_y,\ell_z\}}{r-1},
\qquad
d_{95}^{*}
=
\frac{
\operatorname{HD}_{95}(\mathcal{M}_1,\mathcal{M}_2)
}{
h
},
\label{eq:normalized_hd95}
\end{equation}
where $\ell_x$, $\ell_y$, and $\ell_z$ are the bounding-box dimensions of the reference surface and $r$ is the selected voxel-grid resolution.

\subsubsection{Adaptive stopping and rollback}

For either stage, let $J^{(t)}$ denote the corresponding smoothness objective, $E_{\mathrm{L}}$ for the Laplacian stage or $R(\mathcal{M})$ for the Taubin stage. The relative improvement was
\begin{equation}
I^{(t)}
=
100
\frac{
J^{(t-1)}-J^{(t)}
}{
J^{(t-1)}
}.
\label{eq:relative_smoothing_improvement}
\end{equation}

An iteration was accepted only if the stage-specific objective did not worsen and all geometric-fidelity constraints were satisfied; otherwise, the procedure reverted to the best previously accepted state. The stopping threshold was $0.20\%$ improvement with a patience of two evaluations. Volume/principal-axis limits were $0.6\%/3.0\%$ during the Laplacian stage and $1.0\%/3.0\%$ for the final surface, with normalized HD95 limits of $1.0$ relative to the Laplacian-smoothed surface and $1.2$ relative to the original surface. A maximum of $100$ iterations per stage served only as a safety limit.

\begin{figure}
    \centering
    \includegraphics[width=\linewidth]{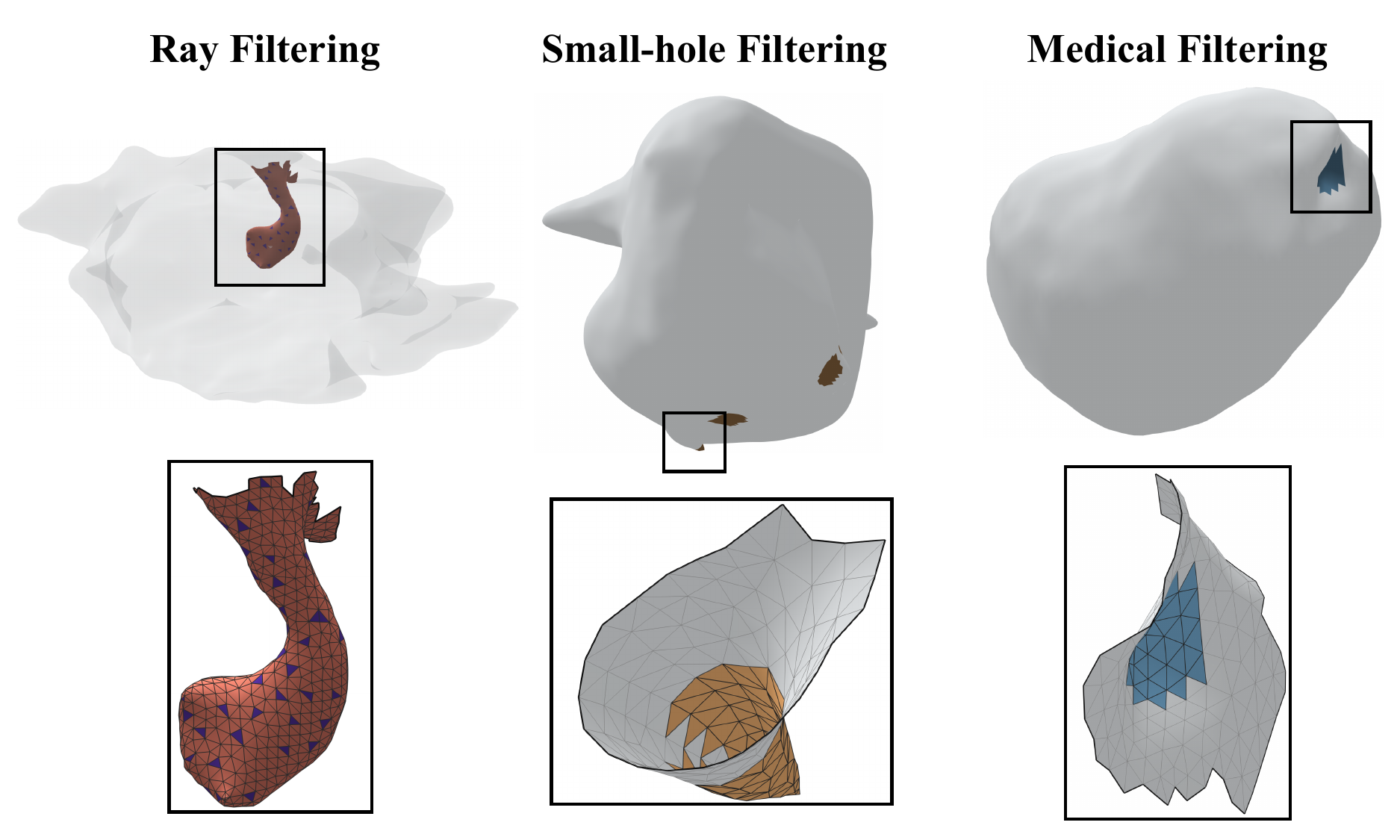}
    \caption{\textbf{Qualitative examples of local artifact detection using
    three complementary filtering strategies.}
    From left to right: ray-based inward-fold detection, local small-hole
    detection, and CT-informed geometric screening. The top row shows the
    location of each detected region on the complete surface, and the bottom
    row shows the corresponding enlarged mesh patch.}
    \label{fig:artifacts_filtering}
\end{figure}

\subsection{Detection of local surface artifacts}
\label{sec:local_artifact_repair}

Although the preceding smoothing and topology correction steps removed major surface irregularities, small localized artifacts could remain without altering the global mesh topology. These defects typically appeared as narrow inward folds, steep-walled pits, or small recessed regions and therefore could not be identified reliably from genus or connectivity alone. We developed a three-stage local detection framework consisting of an initial $\alpha$-shape-based candidate localization, followed by complementary ray-based fold detection, local small-hole detection, and CT-informed geometric screening (Fig.~\ref{fig:artifacts_filtering}). 

\subsubsection{\texorpdfstring{$\alpha$}{alpha}-shape candidate localization}

Let $\mathcal{M}=(V,F)$ denote the processed triangular mesh. A three-dimensional $\alpha$-shape was constructed from the mesh vertices as an external geometric reference~\citep{edelsbrunner1994three}. Rather than using a fixed dimensional value, the $\alpha$ parameter was scaled relative to the critical value required to obtain a single connected $\alpha$-shape region,
\begin{equation}
    \alpha
    =
    1.4\,\alpha_{\mathrm{one}},
    \label{eq:artifact_alpha_value}
\end{equation}
where $\alpha_{\mathrm{one}}$ denotes the corresponding one-region critical value.

For each input vertex $\mathbf{v}_i$, let $d_i$ denote its distance to the $\alpha$-shape boundary and let $\overline{\ell}$ denote the mean mesh-edge length. Vertices not belonging to the $\alpha$-shape boundary were considered candidate vertices when $d_i
    >
    2.5\,\overline{\ell}$.
The initial candidate mask was formed by all faces incident to at least one candidate vertex. Candidate faces were partitioned into edge-connected components and processed independently. The $\alpha$-shape stage was used only to define a sensitive search domain and did not directly determine the faces subsequently removed.

\subsubsection{Ray-based inward-fold detection}
\label{sec:ray_folding_detection}

The first filtering stage targeted narrow inward folds and re-entrant surface regions, as illustrated in the left panel of Fig.~\ref{fig:artifacts_filtering}. A locally regular surface is approximately single-valued relative to its outward direction, whereas an inward fold may produce multiple separated intersections along the same inward-directed ray.

For each candidate component containing at least $10$ faces, three rings of surrounding non-candidate faces were used to estimate a local reference frame from area-weighted face centroids and consistently oriented normals. Let $\widetilde{\ell}_{\mathrm{loc}}$ denote the median edge length within this local region. Parallel rays were sampled over the projected candidate component with spacing $h_{\mathrm{ray}}=1.5\,\widetilde{\ell}_{\mathrm{loc}}$, using at least a $3\times3$ grid and at most $1200$ rays per component. Each ray was cast inward along the negative local reference normal,
\begin{equation}
    \mathbf{x}_r(t)
    =
    \mathbf{o}_r-t\mathbf{n}_k,
    \qquad
    t>0,
    \label{eq:local_inward_ray}
\end{equation}
where $\mathbf{o}_r$ denotes the ray origin.

Ray--triangle intersections within the local region were computed using the two-sided M\"oller--Trumbore algorithm \citep{moller1997fast}. Nearly tangential and numerically coincident intersections were excluded or merged before analysis. If a ray intersected the candidate component more than once, the normalized separation between its first and last candidate crossings was defined as
\begin{equation}
    \Delta_r^{*}
    =
    \frac{s_{r,m_r}-s_{r,1}}
         {\widetilde{\ell}_{\mathrm{loc}}}
    \geq 2.5.
    \label{eq:normalized_ray_span}
\end{equation}

Let $N_{\mathrm{v}}$, $N_{\mathrm{m}}$, and $N_{\mathrm{d}}$ denote the numbers of valid, multi-hit, and deep multi-hit rays, respectively. A component was retained only when $N_{\mathrm{v}}\geq5$, $N_{\mathrm{m}}\geq3$, $N_{\mathrm{m}}/N_{\mathrm{v}}\geq0.05$, $N_{\mathrm{d}}\geq2$, and $N_{\mathrm{d}}/N_{\mathrm{v}}\geq0.02$. These joint count and proportion requirements reduced sensitivity to isolated numerical intersections.

Deep rays were grouped using $8$-neighbor connectivity on the sampling grid, and clusters containing fewer than two deep rays were discarded. Intersected candidate faces from each retained cluster were used as seed faces, while candidate faces adjacent to the component exterior formed a two-ring mouth-anchor band.

A weighted face-adjacency graph was then constructed over the candidate component. Let $d_{\mathrm{s}}(f)$ and $d_{\mathrm{m}}(f)$ denote shortest-path distances from face $f$ to the seed and mouth-anchor sets, respectively. We defined the normalized seed--mouth competition score
\begin{equation}
    q(f)
    =
    \frac{
        d_{\mathrm{m}}(f)
    }{
        d_{\mathrm{s}}(f)+d_{\mathrm{m}}(f)
    }.
    \label{eq:seed_mouth_score}
\end{equation}
Candidate thresholds from $0.90$ to $0.05$ were evaluated in steps of $0.025$. The highest threshold was retained for which the selected region contained all seed faces, formed a single edge-connected patch with one unbranched boundary loop, and occupied no more than $65\%$ of the original candidate component. Components for which no admissible region was found were not automatically included in the removal mask.

Finally, to restrict the mask to sufficiently recessed geometry, inward depth was measured relative to the local reference plane. Faces were retained only when their inward depth was at least $40\%$ of the 90th percentile of positive seed-face depths. Connected regions containing fewer than $20$ faces were discarded. The remaining inward region was expanded by at most four face-adjacency rings within the previously accepted graph-based region, and the largest connected patch was retained as the ray-selected artifact mask.

\subsubsection{Local small-hole detection}

Candidate regions not selected by the ray-based stage were subsequently examined for small, steep-walled pits (middle panel of Fig.~\ref{fig:artifacts_filtering}). This detector combined relative triangle size and shape, adjacent-face normal variation, and a multiscale smoothing response.

For face $f_i$ with area $A_i$, relative area was evaluated with respect to both the surrounding non-candidate surface and a local mesh neighborhood. The effective area ratio was
\begin{equation}
    r_i
    =
    \min
    \left\{
        \frac{
            A_i
        }{
            \operatorname{median}_{j\in\mathcal{S}_i}A_j
        },
        \frac{
            A_i
        }{
            \operatorname{median}_{j\in\mathcal{N}_i}A_j
        }
    \right\},
    \label{eq:effective_face_area_ratio}
\end{equation}
where $\mathcal{S}_i$ denotes the surrounding reference faces and $\mathcal{N}_i$ a local multi-ring neighborhood. Triangle regularity was measured by
\begin{equation}
    q_i
    =
    \frac{
        4\sqrt{3}\,A_i
    }{
        \ell_{i,1}^{2}
        +
        \ell_{i,2}^{2}
        +
        \ell_{i,3}^{2}
    },
    \label{eq:triangle_quality}
\end{equation}
where $\ell_{i,1}, \ell_{i,2}, \ell_{i,3}$ denote the side lengths of face $f_i$; $q_i=1$ for an equilateral triangle. Let $\theta_i$ denote the maximum normal-angle difference between $f_i$ and its edge-adjacent faces. A geometry-based seed was generated when $ \left(
        r_i\leq0.55
        \;\lor\;
        q_i\leq0.40
    \right)
    \land
    \left(
        \theta_i\geq35^{\circ}
    \right)$.

To detect recessed pits not characterized by unusually small or slender triangles, we additionally measured their response to mild implicit cotangent-Laplacian smoothing~\citep{desbrun1999implicit,meyer2003discrete}. For vertex $\mathbf{v}_p$, the normal displacement at smoothing scale $t$ was normalized by the global mean edge length $\overline{\ell}$,
\begin{equation}
    s_p^{(t)}
    =
    \frac{
        \left(
            \mathbf{v}_p^{(t)}
            -
            \mathbf{v}_p^{(0)}
        \right)^{\mathsf T}
        \mathbf{n}_p
    }{
        \overline{\ell}
    },
    \qquad
    t\in\{1,2,4\}.
    \label{eq:multiscale_inward_response}
\end{equation}
The face response was the maximum value across its vertices and the three smoothing scales. A smoothing-response seed was retained when the response exceeded both $0.08$ edge lengths and the local 95th percentile.

The union of the geometry- and smoothing-based seeds was expanded by three face rings within the local candidate domain. Enclosed internal islands were filled to obtain spatially continuous patches. A patch was retained only if its seed fraction was at least $0.02$, it contained no more than $220$ faces, and it occupied no more than $20\%$ of its local growth domain.

\subsubsection{CT-informed geometric screening}

A third complementary stage screened residual candidate regions using geometric patterns motivated by common CT-derived surface artifacts, as described in the right panel of Fig.~\ref{fig:artifacts_filtering}. Four evidence branches were considered: thin opposing surface sheets, local inward recession, risk of opposing-wall merging, and staircase or terrace patterns associated with the through-plane direction~\citep{CEVIDANES2010361,Ito2019,Shirshin2021,Moench2011}.

Let $h$ denote the effective voxel scale associated with the reconstructed surface. Thin-wall evidence was recorded when the estimated distance $t_i$ between approximately opposing surface sheets satisfied $t_i/h \leq 2.5$.
Local recession was defined relative to a fitted reference plane and required an inward depth of at least $0.5h$ together with local spatial continuity.

A wall-break score was defined by
\begin{equation}
    b_i
    =
    \frac{
        e_i+e_i^{\mathrm{opp}}
    }{
        t_i
    },
    \label{eq:wall_break_risk}
\end{equation}
where $e_i$ and $e_i^{\mathrm{opp}}$ denote multiscale surface-deviation responses on the two opposing sheets. Faces with $b_i\geq0.70$ provided wall-break evidence. Staircase evidence was derived from local variation in face orientation relative to the prescribed stack direction and from small, nearly planar terraces.

To reduce false-positive removal of anatomically plausible concavities, staircase evidence was not sufficient by itself. A strict seed required at least two evidence branches, with at least one corresponding to thin-wall, recession, or wall-break evidence. Connected support regions were subsequently formed around these multi-evidence seeds.

\begin{figure}
    \centering
    \includegraphics[width=\linewidth]{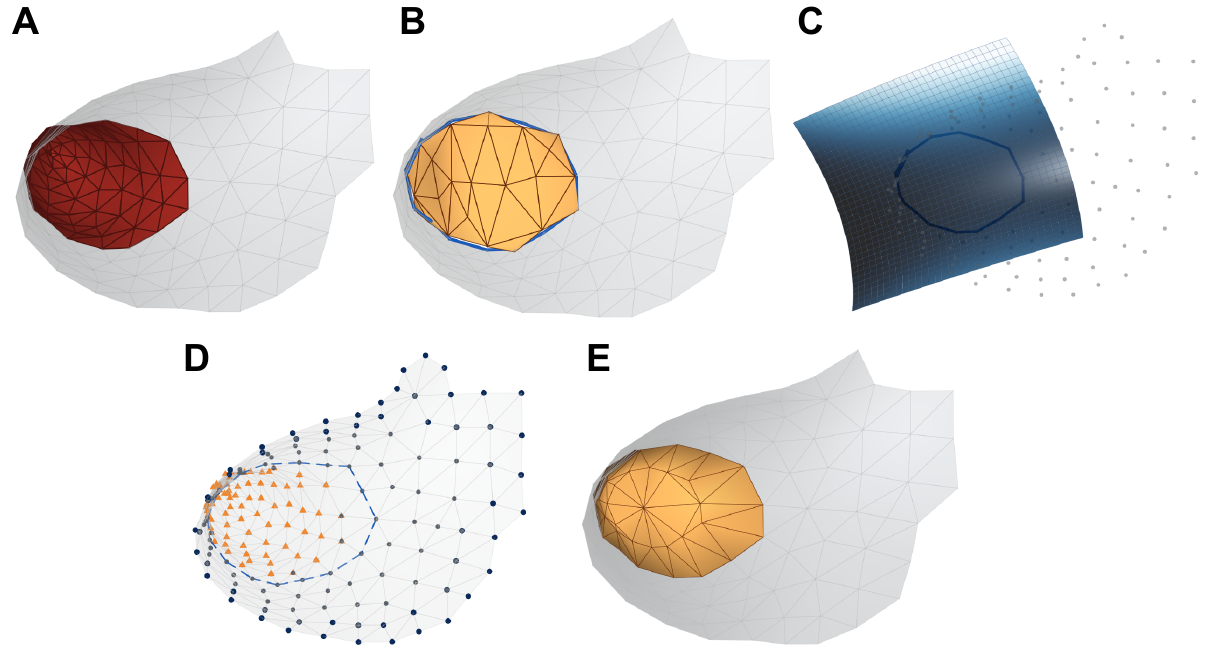}
    \caption{\textbf{Local artifact excision and surface-reconstruction workflow.} \textbf{(A)} Artifact excision. \textbf{(B)} Initial triangulation and conforming subdivision. \textbf{(C)} Quadratic reference-surface estimation. \textbf{(D)} Constrained cotangent bi-Laplacian fairing. \textbf{(E)}~Final reconstructed surface after local Taubin smoothing.}
    \label{fig:artifacts_repairing}
\end{figure}

\subsection{Repair and reconstruction of local surface artifacts}
\label{sec:local_excision_reconstruction}

Let $\mathcal{C}_{\mathrm{ray}}$, $\mathcal{C}_{\mathrm{hole}}$, and $\mathcal{C}_{\mathrm{CT}}$ denote the faces selected by the three detection stages. Their union
\begin{equation}
\mathcal{C}
=
\mathcal{C}_{\mathrm{ray}}
\cup
\mathcal{C}_{\mathrm{hole}}
\cup
\mathcal{C}_{\mathrm{CT}}
\end{equation}
defined the candidate removal mask passed to the reconstruction stage. The local repair procedure consisted of artifact excision, boundary-aware triangulation, conforming subdivision, quadratic reference-surface estimation, constrained cotangent bi-Laplacian fairing, and final local Taubin smoothing. The complete repair sequence is summarized in Fig.~\ref{fig:artifacts_repairing}.

\subsubsection{Local excision and initial patch construction}

The selected artifact faces were removed from the source mesh, the unreferenced vertices were discarded, and the local boundary loops were extracted from the edges incident to exactly one retained face. This excision step corresponds to Fig.~\ref{fig:artifacts_repairing}A. Hole filling followed the standard boundary-triangulation paradigm used in mesh repair~\citep{liepa2003filling}.

For each boundary loop, a best-fitting plane was estimated by singular value decomposition. When the planar projection produced a valid non-self-intersecting polygon, the opening was filled using constrained Delaunay triangulation~\citep{shewchuk1996triangle}. If planar triangulation was not admissible because of geometric or connectivity conflicts, a three-dimensional conflict-aware ear-clipping procedure was used instead. A centroid fan was retained only as a final fallback. New triangles were oriented consistently with the retained surface along the shared boundary. The initial patch obtained after this step is presented in Fig.~\ref{fig:artifacts_repairing}B.

To provide sufficient degrees of freedom for subsequent geometric refinement, long edges in the reconstructed patch were subdivided conformingly. Let $\ell_{\mathrm{med}}$ denote the median edge length of the input mesh. Edges satisfying $\ell_e > 1.5\,\ell_{\mathrm{med}}$
were split at their midpoints, for at most three subdivision passes.

\subsubsection{Quadratic local reference surface}

The initial triangulation determines patch connectivity but does not by itself recover the local surface geometry. We therefore estimated a smooth reference surface from five rings of measured faces surrounding each reconstructed patch. The resulting quadratic reference-surface construction is represented in Fig.~\ref{fig:artifacts_repairing}C.

For patch $\mathcal{P}_k$, an orthonormal local frame $\{\mathbf{b}_{1,k},\mathbf{b}_{2,k},\mathbf{n}_k\}$ was estimated from the surrounding measured vertices. In local coordinates $(u,v,h)$, the reference geometry was represented by the quadratic height field
\begin{equation}
    \widehat{h}_k(u,v)
    =
    \boldsymbol{\phi}
    \left(
        \frac{u}{s_k},
        \frac{v}{s_k}
    \right)^{\mathsf T}
    \mathbf{a}_k,
    \label{eq:quadratic_height_field}
\end{equation}
where $    \boldsymbol{\phi}(\widetilde{u},\widetilde{v})
    =
    \begin{bmatrix}
        \widetilde{u}^{2} &
        \widetilde{u}\widetilde{v} &
        \widetilde{v}^{2} &
        \widetilde{u} &
        \widetilde{v} &
        1
    \end{bmatrix}^{\mathsf T}$.
The coefficient vector was obtained by distance-weighted, weakly regularized least squares,
\begin{equation}
    \mathbf{a}_k
    =
    \underset{\mathbf{a}\in\mathbb{R}^{6}}{\arg\min}
    \left[
        \sum_{i\in\mathcal{V}^{\mathrm{ref}}_k}
        \omega_i
        \left(
            \boldsymbol{\phi}_i^{\mathsf T}\mathbf{a}
            -
            h_i
        \right)^2
        +
        \varepsilon_{\mathrm{q}}
        \|\mathbf{a}\|_2^2
    \right],
    \label{eq:quadratic_weighted_fit}
\end{equation}
with larger weights assigned to measured vertices closer to the reconstructed region. The fit was omitted when fewer than $12$ measured reference vertices were available or when the local frame was numerically degenerate.

Only interior patch vertices were projected toward the fitted reference surface; boundary vertices remained fixed. The normal displacement of each interior vertex was clipped to $|d_i| \leq 2\,\ell_{\mathrm{med}}$,
thereby limiting the effect of an unstable local fit.

\subsubsection{Constrained cotangent bi-Laplacian fairing}

Following reference-surface estimation, the reconstructed patch and four surrounding face rings were refined jointly. Cotangent weights were used to construct a discrete Laplace--Beltrami operator~\citep{meyer2003discrete}. In this subsection, the positive-semidefinite sign convention was used. Vertices on the outer boundary of the local fairing region were fixed, providing Dirichlet boundary conditions and preventing repair-induced deformation from propagating into the distant measured surface. The constrained fairing stage is depicted in Fig.~\ref{fig:artifacts_repairing}D.

Let $\mathcal{U}_k$ and $\mathcal{D}_k$ denote the movable and fixed vertices, respectively, and partition the relevant Laplacian rows as $    L_{\Omega_k}
    =
    \begin{bmatrix}
        A_k & B_k
    \end{bmatrix}$.
The movable vertex positions were obtained by minimizing
\begin{equation}
\resizebox{.9\linewidth}{!}{$
\begin{aligned}
    E_k(\mathbf{X}_{\mathcal{U}_k})
    ={}&
    \left\|
        A_k\mathbf{X}_{\mathcal{U}_k}
        +
        B_k\mathbf{X}_{\mathcal{D}_k}
    \right\|_F^2
    +
    \sum_{i\in\mathcal{U}_k}
    \gamma_i
    \left\|
        \mathbf{x}_i-\mathbf{x}^{\mathrm{ref}}_i
    \right\|_2^2
    +
    \varepsilon_{\mathrm{f}}
    \left\|
        \mathbf{X}_{\mathcal{U}_k}
    \right\|_F^2.
\end{aligned}
$}
\label{eq:repair_bilaplacian_energy}
\end{equation}
This formulation combines Laplacian fairing~\citep{desbrun1999implicit,meyer2003discrete} with soft positional constraints. The positional weights were
\begin{equation}
    \gamma_i
    =
    \begin{cases}
        0.5,
        & i \text{ belongs to the reconstructed patch},\\
        10,
        & i \text{ belongs to the measured surroundings}.
    \end{cases}
    \label{eq:repair_positional_weights}
\end{equation}
Thus, reconstructed vertices were allowed to adapt to the inferred local geometry, whereas measured surrounding vertices were strongly constrained to their original positions. The numerical regularization factor was $10^{-8}$ times the mean diagonal magnitude of $A_k^{\mathsf T}A_k$.

\subsubsection{Local smoothing and post-repair validation}

After bi-Laplacian fairing, the reconstructed patch and five surrounding face rings underwent five iterations of local Taubin smoothing~\citep{taubin1995signal}, with $\lambda_{\mathrm{R}}=0.25$ and $\mu_{\mathrm{R}}=-0.26$.
Vertices on the outer boundary of the local smoothing region remained fixed, ultimately yielding the final locally smoothed patch shown in Fig.~\ref{fig:artifacts_repairing}E.

To prevent local element inversion during geometric refinement, quadratic projection, bi-Laplacian fairing, and each Taubin half-step were subjected to backtracking. Proposed displacements were progressively reduced until every affected triangle retained positive area and preserved its orientation relative to the previous state. If $    \mathbf{a}_f
    =
    \left(
        \mathbf{x}_{f_2}-\mathbf{x}_{f_1}
    \right)
    \times
    \left(
        \mathbf{x}_{f_3}-\mathbf{x}_{f_1}
    \right)$
denotes the oriented area vector of face $f$, an update was accepted only if
\begin{equation}
    \left\|
        \mathbf{a}^{\mathrm{new}}_f
    \right\|_2
    >
    \varepsilon_{\mathrm{A}},
    \qquad
    \left(
        \mathbf{a}^{\mathrm{old}}_f
    \right)^{\mathsf T}
    \mathbf{a}^{\mathrm{new}}_f
    >
    0
    \label{eq:repair_update_safety}
\end{equation}
for all affected faces. Updates for which no admissible displacement scale could be found were rejected.

After reconstruction, vertex-touching floating fragments were removed by retaining the largest edge-connected surface component, and the mesh indices were compacted. The complete strict topology and mesh-integrity validation defined in Section~\ref{sec:adaptive_voxelization} was then repeated. Altogether, the pipeline produces connected, closed, consistently oriented genus-$0$ meshes without boundary edges, non-manifold edges, orientation conflicts, duplicate faces, degenerate faces, or unreferenced vertices for the subsequent spherical parameterization and spherical harmonic analysis.

\section{Shape analysis of lymph node surfaces}
\label{sec:shape_analysis}

\subsection{Spherical mapping}
\label{sec:spherical_mapping}

After geometric processing and topology validation, each genus-$0$ lymph node surface was parameterized onto the unit sphere to provide a common domain for subsequent spherical harmonic analysis. Three candidate parameterizations were considered: a spherical conformal map followed by M\"obius area correction~\citep{choi2015flash,choi2020partialwelding}, an area-preserving spherical density-equalizing map initialized from the conformal parameterization~\citep{lyu2024spherical}, and a spherical Tutte map~\citep{tutte1963draw} included as a robust alternative.

Each candidate mapping was subjected to validity screening according to its parameterization characteristics. For the conformal and area-preserving parameterizations, a mapping was considered numerically valid only when the mapped surface exhibited consistent local face orientation, contained no degenerate spherical triangles or collapsed mapped vertices, and had a total spherical triangle area within $5\%$ of $4\pi$. When geometry-related validity issues occurred in the conformal or area-preserving parameterizations, the spherical Tutte mapping was used as a fallback. 

Mapping selection was performed independently for each lymph node and each SH reconstruction degree (the SH reconstruction procedures are explained in the following section). For a given degree $L$, the SH-reconstructed surface was obtained using each admissible spherical mapping, and agreement with the processed reference surface was quantified using the normalized symmetric bidirectional nearest-neighbor root-mean-square distance
\begin{equation} 
\resizebox{.9\linewidth}{!}{$
E_{\mathrm{RMS}}
=
\frac{1}{D_{\mathrm{bbox}}}
\sqrt{
\frac{1}{2}
\left[
\frac{1}{N}\sum_{i=1}^{N}
\min_j
\|\mathbf{x}_i-\widehat{\mathbf{x}}_j\|_2^2
+
\frac{1}{\widehat{N}}\sum_{j=1}^{\widehat{N}}
\min_i
\|\widehat{\mathbf{x}}_j-\mathbf{x}_i\|_2^2
\right]
},
$}
\label{eq:mapping_selection_rms}
\end{equation}
where $\mathbf{x}_i$ and $\widehat{\mathbf{x}}_j$ denote vertices of the processed reference and SH-reconstructed surfaces, respectively, and $D_{\mathrm{bbox}}$ is the diagonal length of the reference-surface bounding box. Among the admissible mappings, the mapping yielding the smallest $E_{\mathrm{RMS}}$ was retained for that lymph node and SH degree. Symmetric mean nearest-neighbor distance and HD95 were additionally recorded for quality assessment but were not used for mapping selection. The entire mapping-selection procedure was based solely on geometric criteria and did not use metastatic status, class labels, or predictive outcomes.

\subsection{Spherical harmonic representation}
\label{sec:spherical_harmonic_representation}

Using the degree-specific spherical parameterization selected above, each lymph node surface was represented using real spherical harmonics (SH), which provide a hierarchical spectral representation of genus-$0$ three-dimensional surfaces~\citep{brechbuehler1995parametrization,styner2006spharm,kazhdan2003}. Let
$\mathbf{x}(\theta,\phi)
=
\begin{bmatrix}
x(\theta,\phi)\\
y(\theta,\phi)\\
z(\theta,\phi)
\end{bmatrix}$
denote the Cartesian surface coordinates associated with spherical coordinates $(\theta,\phi)$. The degree-$L$ approximation was
\begin{equation}
\mathbf{x}_{L}(\theta,\phi)
=
\sum_{\ell=0}^{L}
\sum_{m=-\ell}^{\ell}
\mathbf{c}_{\ell m}
Y_{\ell}^{m}(\theta,\phi),
\label{eq:sh_surface_representation}
\end{equation}
where $Y_{\ell}^{m}$ denotes a real orthonormal spherical harmonic basis function of degree $\ell$ and order $m$, $\mathbf{c}_{\ell m}\in\mathbb{R}^{3}$ is the corresponding coefficient vector, and $L$ is the maximum harmonic degree.

In the numerical implementation, SH coefficients were estimated after weighted centroid removal using spherical triangle areas as quadrature weights. This weighting reduces sensitivity to nonuniform vertex sampling on the spherical parameterization. The weighted centroid was restored after reconstruction.

For the primary lymph node analyses, we evaluated $L\in\{5,8,10,15,20,34\}$, and a degree-$L$ expansion contains $N_{\mathrm{SH}}=(L+1)^2$
three-dimensional coefficient vectors. The corresponding reconstructions were subsequently used for geometric feature extraction, multi-resolution analysis, and evaluation of geometric fidelity and predictive performance. A separate degree range was used for the LIDC-IDRI cross-organ experiment, as specified in Section~\ref{sec:lidc_crossorgan_method}.

\subsection{Surface feature extraction}
\label{sec:surface_feature_extraction}

To characterize lymph node morphology beyond conventional global shape measurements, geometric descriptors were extracted from the processed reference surfaces and from the SH-reconstructed surfaces. For the multi-resolution analysis, identical feature-extraction procedures and numerical settings were applied independently to the SH reconstructions at different degrees.

\subsubsection{Conventional geometric shape features}
\label{sec:conventional_shape_features}

Conventional global morphology was characterized using the 14 three-dimensional shape features defined in the PyRadiomics framework~\citep{vangriethuysen2017pyradiomics}. The same descriptors were extracted from the original, processed-reference, and SH-reconstructed surfaces. These features, denoted \textit{Py14}, also served as the conventional global-morphology baseline.

\subsubsection{Extended surface descriptors and feature-family organization}
\label{sec:extended_surface_features}

Extended descriptors were extracted to characterize complementary aspects of global geometry, local differential geometry, intrinsic spectral geometry, shape distributions, symmetry, and SH spectral organization. The descriptors were organized \textit{a priori} into 20 geometrically defined feature families (Table~\ref{tab:feature_family_organization}) using established formulations for differential geometry, intrinsic spectral descriptors, geodesic measures, shape distributions, and symmetry \citep{meyer2003discrete,koenderink1992surface,reuter2006shapedna,sun2009hks,aubry2011wks,crane2013geodesics,osada2001,kazhdan2004}. The extended set comprised 352 variables; together with \textit{Py14}, this yielded 366 candidate variables for multi-resolution analysis.

\begin{table}[!t]
\centering
\footnotesize
\setlength{\tabcolsep}{2.5pt}
\renewcommand{\arraystretch}{0.95}
\caption{\textbf{Feature families used in the multi-resolution SH analysis.}}
\label{tab:feature_family_organization}
\begin{tabularx}{\linewidth}{>{\raggedright\arraybackslash}p{0.4\linewidth}>{\raggedright\arraybackslash}X>{\centering\arraybackslash}p{0.05\linewidth}}
\hline
Feature family & Main descriptors & $n$ \\
\hline
Convexity & Convex-hull measures & 6 \\
Radial morphology & Radial-distance statistics & 15 \\
Reflection symmetry & Principal-plane reflection errors & 9 \\
Mean curvature & Mean-curvature statistics & 26 \\
Gaussian curvature & Gaussian-curvature statistics & 13 \\
Shape Index and Curvedness & Local shape and bending measures & 28 \\
Curvature integrals & Integrated curvature measures & 9 \\
Curvature regions & Curvature-region descriptors & 16 \\
Laplace--Beltrami spectrum & Spectral eigenvalue descriptors & 45 \\
Heat Kernel Signature & Multi-scale HKS descriptors & 60 \\
Wave Kernel Signature & Multi-scale WKS descriptors & 50 \\
Geodesic geometry & Intrinsic extent measures & 3 \\
$D2$ shape distribution & Pairwise-distance distribution & 10 \\
$D3$ shape distribution & Triangle-area distribution & 10 \\
$D4$ shape distribution & Tetrahedral-volume distribution & 10 \\
Spherical extent & Directional extent and anisotropy & 12 \\
Central asymmetry & Opposite-direction asymmetry & 10 \\
$180^\circ$ rotational symmetry & Rotational discrepancy measures & 8 \\
\textit{Py14} & Conventional PyRadiomics shape features & 14 \\
SH spectral energy & SH energy-distribution descriptors & 12 \\
\hline
\textbf{Total} & & \textbf{366} \\
\hline
\end{tabularx}
\end{table}

\subsection{Quantitative evaluation and statistical analysis}
\label{sec:quantitative_evaluation}

Quantitative analyses evaluated geometric validity, multi-resolution representation, predictive performance, perturbation robustness, and transportability across independent datasets. The complete set of $1,769$ surfaces was used for geometric processing and SH analyses, while predefined subsets were used for computationally intensive paired evaluations as specified below. The independent multicenter lymph node cohort and the LIDC-IDRI lung-nodule cohort were analyzed separately from the primary cohort for label-free external replication and cross-organ validation, respectively.

The individual lymph node was the unit of prediction and performance evaluation, whereas the patient was the grouping and resampling unit used to prevent information leakage and account for within-patient dependence. For all predictive analyses, outer and inner cross-validation were performed at the patient level, with all lymph nodes from the same patient assigned to the same fold. Data-dependent preprocessing and hyperparameter selection used training data only, and identical outer partitions were retained for paired model comparisons. For descriptive interpretation of metastasis-associated geometric phenotypes, univariate AUC and Cliff's $\delta$ were additionally used to summarize group separation between metastatic and non-metastatic lymph nodes. 

\subsubsection{Topological validity and downstream computational quality}
\label{sec:topological_computational_quality}

Topological validity was evaluated on all $1,769$ surfaces before and after geometric processing using the strict validation criteria defined in Section~\ref{sec:adaptive_voxelization}. The primary endpoint was strict topology validity, which required a single connected, closed, consistently oriented genus-$0$ surface without boundary edges, non-manifold elements, orientation conflicts, duplicate faces, degenerate faces, or unreferenced vertices. Single-component structure, genus-$0$ topology, and outward orientation were additionally summarized as representative component criteria.

Downstream computational quality was evaluated on a randomly selected fixed subset of $500$ paired original and processed surfaces using two complementary measures: triangle quality and adjacent-face normal variation. Triangle quality was quantified using the normalized area-based metric defined in Eq.~\eqref{eq:triangle_quality}, where values closer to one indicate more equilateral triangles. The reported metric, triangle quality (5th percentile), was defined as the fifth percentile of $q_i$ over all mesh faces. Adjacent-face normal variation, quantified by the normal jump (95th percentile, degrees), was defined as the area-weighted 95th percentile of dihedral angles between neighboring face normals. Higher triangle quality and lower normal jump indicate improved numerical suitability for subsequent differential-geometric computation.

\subsubsection{Univariate discrimination of conventional shape descriptors}
\label{sec:py14_univariate_methods}

To assess whether SH reconstruction altered the discriminative behavior of conventional morphology, each \textit{Py14} descriptor was evaluated independently on the original surface and on the six SH reconstructions using a single-feature L2-regularized logistic-regression model. This analysis used 10 repeats of five-fold patient-grouped stratified outer cross-validation, with five-fold patient-grouped inner cross-validation for regularization selection. Identical outer splits were used across descriptors and surface representations, and out-of-fold probabilities were averaged across the 10 repeats before calculation of the final AUC.

For descriptor $f$ and degree $L$, the change relative to the original surface was
\begin{equation}
\Delta\mathrm{AUC}_{f,L}
=
\mathrm{AUC}_{f,L}^{\mathrm{SH}}
-
\mathrm{AUC}_{f}^{\mathrm{original}}.
\label{eq:py14_delta_auc}
\end{equation}

\subsubsection{Degree-wise spectral localization of discriminative information}
\label{sec:degree_wise_spectral_discrimination}

To characterize the discriminative information carried by different harmonic degrees, we performed a degree-wise spectral analysis using the coefficients from the degree-34 SH representation. This approach avoids reconstructing separate surfaces at each harmonic degree and directly evaluates the contribution of individual degrees within a common high-resolution representation~\citep{kazhdan2003}. For degree $\ell$, the SH energy was defined as
\begin{equation}
E_{\ell}
=
\sum_{m=-\ell}^{\ell}
\left\|
\mathbf{c}_{\ell m}
\right\|_2^2,
\end{equation}
and its relative contribution to nonconstant shape variation was
\begin{equation}
P_{\ell}
= E_{\ell} \left/
\left(
\displaystyle\sum_{k=1}^{34}E_k\right)\right.,
\qquad
\ell=1,\ldots,34.
\label{eq:relative_degree_energy}
\end{equation}
The degree-$0$ term was excluded because it represents the constant component and was not treated as shape variation.

The univariate discriminative strength of each degree was summarized using the folded AUC
\begin{equation}
\mathrm{AUC}^{*}_{\ell}
=
\max
\left\{
\mathrm{AUC}_{\ell},
1-\mathrm{AUC}_{\ell}
\right\},
\label{eq:degree_oriented_auc}
\end{equation}
for which $0.5$ represents chance-level discrimination and which summarizes discrimination irrespective of the direction of association. Degree-wise energy and discriminative performance were examined jointly to distinguish the amount of geometric information represented at a given spectral scale from its relevance to metastatic-status discrimination.

\subsubsection{Geometric fidelity and high-resolution representation efficiency}
\label{sec:sh_geometric_fidelity}

Direct reconstruction fidelity was evaluated on $500$ lymph-node surfaces with complete measurements at all six SH degrees. For each reconstruction, four complementary geometric errors were calculated relative to the processed reference surface: normalized symmetric mean nearest-neighbor distance, normalized symmetric HD95, relative surface-area error, and relative volume error.

Because these measures have different numerical scales, each error was converted to a pooled empirical fidelity score. Let $e_{iLq}$ denote geometric error metric $q$ for lymph node $i$ reconstructed at degree $L$, and let $r_{iLq}$ denote its rank among all lymph-node--degree observations for that metric, with smaller errors assigned lower ranks. The corresponding fidelity score was
\begin{equation}
F_{iLq}
=
1-\frac{r_{iLq}-1}{N_{\mathrm{pool}}-1},
\label{eq:metric_fidelity_score}
\end{equation}
where $N_{\mathrm{pool}}=3000$ for the $500$ lymph-node surfaces evaluated at six degrees. The composite geometric fidelity score was defined as
\begin{equation}
F_{iL}
=
\frac{1}{4}\sum_{q=1}^{4}F_{iLq}.
\label{eq:composite_geometric_fidelity}
\end{equation}
Larger values indicate better overall agreement with the processed reference geometry. The score is a cohort-relative summary and should not be interpreted as an absolute physical measure of reconstruction error. Degree-specific medians and 95\% confidence intervals were estimated using $5,000$ bootstrap resamples.

Because predictive optimization and faithful geometric representation need not favor the same SH degree, degree $34$ was additionally examined as the high-resolution representation. Preservation of conventional morphology was assessed using Pearson correlation, Spearman correlation, and Lin's concordance correlation coefficient across the 14 PyRadiomics shape features. Representation efficiency was evaluated by comparing the $1,225$ three-dimensional SH coefficient vectors at degree $34$ with the number of vertices in the processed reference meshes. This comparison was interpreted as a reduction in subject-specific geometric representation elements rather than as a direct estimate of runtime or byte-level storage efficiency.

\subsubsection{Fixed-resolution and family-specific mixed-resolution prediction}
\label{sec:mixed_resolution_modeling}

Predictive modeling was performed on all $1,769$ lymph nodes from $214$ patients. At each fixed SH degree, all $366$ candidate variables from the 20 predefined feature families were used to construct a fixed-resolution model. A family-specific mixed-resolution model was constructed to allow each feature family to select its preferred SH degree using training data only.

All headline models used three repeats of five-fold patient-grouped stratified outer cross-validation, with identical outer partitions locked across all SH degrees and model comparisons. Within each outer training set, five-fold patient-grouped stratified inner cross-validation was used for degree selection and regularization tuning. The modeling pipeline consisted of median imputation with missing-value indicators, feature standardization, and class-balanced L2-regularized logistic regression. The regularization parameter was selected from $C\in\{0.01,0.1,1,10,100\}$ according to mean inner-cross-validation AUC.

For the mixed-resolution model, degree selection was performed separately for each of the 20 feature families within each outer training set. For family $j$, models were evaluated at $L\in\{5,8,10,15,20,34\}$, and the degree producing the highest mean inner-cross-validation AUC was selected. Effectively exact ties, defined by a tolerance of $10^{-12}$, were resolved in favor of the lower degree. The feature blocks corresponding to the 20 selected family-specific degrees were then concatenated, after which the regularization parameter of the combined model was tuned again using inner cross-validation. The held-out outer fold was used only for final prediction. Fixed-resolution comparators used the same 366 variables, modeling pipeline, and locked outer-fold assignments, differing only in that every feature family was represented at the same SH degree.

For each lymph node, the out-of-fold probabilities obtained from the three outer repeats were averaged before calculation of the final node-level AUC. Confidence intervals and paired model differences were estimated using 2,000 stratified patient-cluster bootstrap resamples, with patients serving as the resampling units and all lymph nodes from each selected patient retained jointly.

The original-surface \textit{Py14} model was evaluated on 1,769 lymph nodes from 214 patients as the conventional global-morphology baseline. Its lymph-node-level out-of-fold predictions were compared with those of the mixed-resolution model using paired bootstrap resampling.

\subsection{Resolution-dependent perturbation robustness}
\label{subsec:perturbation_robustness}

To investigate how SH resolution affects sensitivity to small geometric variation, we performed a controlled perturbation analysis at geometric, harmonic, and predictive levels. A single perturbation realization was generated for each processed lymph node surface and propagated through all six SH reconstruction degrees: $L\in\{5,8,10,15,20,34\}$. Complete paired data were available for a randomly selected subset of $324$ lymph-node surfaces throughout the robustness analysis. Each degree retained its own degree-specific spherical parameterization selected from the corresponding clean surface; therefore, between-degree differences reflect the combined behavior of the lower- or higher-resolution SH representation and its clean-selected mapping, rather than spectral truncation alone.

\subsubsection{Controlled geometric perturbation}
\label{sec:controlled_geometric_perturbation}

Let $\mathbf{v}_i$ and $\mathbf{n}_i$ denote the position and outward unit normal of vertex $i$, respectively, and let $(s_x,s_y,s_z)$ denote the voxel spacing in millimetres. To account for anisotropic image resolution, the physical scale corresponding to a one-voxel displacement along the local surface normal was defined as
\begin{equation}
h_i
=
\left[
\left(\frac{n_{i,x}}{s_x}\right)^2
+
\left(\frac{n_{i,y}}{s_y}\right)^2
+
\left(\frac{n_{i,z}}{s_z}\right)^2
\right]^{-1/2}.
\label{eq:normal_voxel_scale}
\end{equation}

A spatially correlated scalar perturbation field was generated from independent standard-normal samples $\boldsymbol{\varepsilon}$ using cotangent finite-element heat diffusion,
\begin{equation}
(\mathbf{M}+t\mathbf{L})\mathbf{z}_{0}
=
\mathbf{M}\boldsymbol{\varepsilon},
\qquad
t=\frac{\rho^2}{2},
\label{eq:perturbation_heat_field}
\end{equation}
where $\mathbf{M}$ is the lumped mass matrix and $\mathbf{L}$ is the positive-semidefinite cotangent Laplacian. The correlation length was fixed a priori at $\rho_{\mathrm{vox}}=1.75$ voxels for all lymph-node surfaces. The corresponding physical correlation length for each case was defined as
\begin{equation}
\rho
=
\rho_{\mathrm{vox}}
(s_xs_ys_z)^{1/3}.
\label{eq:physical_correlation_length}
\end{equation}

The diffused field was standardized using the lumped mass weights to have zero weighted mean and unit weighted root-mean-square magnitude. Extreme field values were then symmetrically clipped at $\pm2.5$ standard deviations, followed by a second weighted standardization.

The normal displacement at vertex $i$ was then defined as
\begin{equation}
d_i
=
\alpha h_i z_i,
\label{eq:perturbation_normal_displacement}
\end{equation}
and the perturbed vertex position was
\begin{equation}
\mathbf{v}^{\mathrm{pert}}_i
=
\mathbf{v}_i
+
d_i\mathbf{n}_i.
\label{eq:perturbed_vertex}
\end{equation}
The perturbation amplitude was fixed at $\alpha=0.2$. The same standardized random field was used for all SH degrees of a given lymph node, thereby preventing degree-specific random perturbations from confounding the comparison. Perturbed surfaces retained the connectivity and vertex correspondence of the corresponding clean surfaces.

Candidate perturbations were accepted only when all affected triangles retained positive area according to the numerical validity criterion used in the implementation and no face-orientation reversal occurred. If these conditions were violated, the candidate was discarded and a new seeded realization was generated.

For each clean--perturbed pair, the spherical parameterization selected from the clean surface was kept fixed and directly applied to the perturbed surface without re-estimation. Therefore, differences between the clean and perturbed representations within each SH degree reflected only the effect of the imposed geometric perturbation, rather than changes caused by remapping. Comparisons across SH degrees used the corresponding degree-specific mappings selected from the clean surfaces, consistent with the main analysis.

\subsubsection{Geometric perturbation transmission across SH degrees}
\label{sec:geometric_perturbation_transmission}

Geometric perturbation transmission was quantified by comparing the RMS displacement before and after SH reconstruction at corresponding spherical sample locations.

For degree $L$, the perturbation transmission ratio was defined as
\begin{equation}
T_L
=
\frac{
\operatorname{RMS}
\left(
\mathbf{x}^{\mathrm{pert}}_{L}
-
\mathbf{x}^{\mathrm{clean}}_{L}
\right)
}{
\operatorname{RMS}
\left(
\mathbf{x}^{\mathrm{pert}}
-
\mathbf{x}^{\mathrm{clean}}
\right)
},
\label{eq:perturbation_transmission}
\end{equation}
where $\mathbf{x}^{\mathrm{clean}}$ and $\mathbf{x}^{\mathrm{pert}}$ denote corresponding points on the clean and perturbed input surfaces, and the subscript $L$ denotes their SH reconstructions at degree $L$.

Thus, $T_L=1$ indicates complete transmission of the imposed perturbation, whereas $T_L<1$ indicates attenuation.

\subsubsection{Harmonic-domain perturbation analysis}
\label{sec:harmonic_perturbation_analysis}

To characterize the spectral origin of resolution-dependent perturbation transmission, the degree-$34$ SH representation was used as a common high-resolution reference. A single clean-selected degree-$34$ spherical parameterization was used for both the clean and perturbed surfaces so that the spectral comparison was not confounded by differences in spherical mapping. Clean and perturbed surfaces were fitted independently using the fixed parameterization and identical spherical-area quadrature weights.

Let $\mathbf{c}^{\mathrm{clean}}_{\ell m}$ and $\mathbf{c}^{\mathrm{pert}}_{\ell m}$ denote the clean and perturbed SH coefficient vectors at harmonic degree $\ell$ and order $m$. Their coefficient-space difference was
\begin{equation}
\Delta\mathbf{c}_{\ell m}
=
\mathbf{c}^{\mathrm{pert}}_{\ell m}
-
\mathbf{c}^{\mathrm{clean}}_{\ell m},
\end{equation}
and the perturbation energy at degree $\ell$ was defined as
\begin{equation}
E_{\ell}^{\Delta}
=
\sum_{m=-\ell}^{\ell}
\left\|
\Delta\mathbf{c}_{\ell m}
\right\|_2^2.
\label{eq:perturbation_degree_energy}
\end{equation}

Before calculation of spectral energies, SH coefficients were normalized by the equivalent-radius scale of the corresponding clean surface to reduce differences arising solely from overall lymph node size. Degree $0$ was excluded because it represents the constant translation component rather than shape variation.

The relative contribution of degree $\ell$ to the total perturbation spectrum was
\begin{equation}
P_{\ell}^{\Delta}
=
E_{\ell}^{\Delta}\left/
\left(
\displaystyle\sum_{k=1}^{34}
E_k^{\Delta}
\right)\right.,
\qquad
\ell=1,\ldots,34.
\label{eq:relative_perturbation_power}
\end{equation}

The cumulative fraction of degree-$34$ perturbation energy retained after truncation at degree $L$ was defined as
\begin{equation}
R_L
=
\left(\displaystyle\sum_{\ell=1}^{L}
E_{\ell}^{\Delta}\right) \left/
\left(\displaystyle\sum_{\ell=1}^{34}
E_{\ell}^{\Delta}\right)\right..
\label{eq:retained_perturbation_power}
\end{equation}
Accordingly, $1-R_L$ represents the fraction of coefficient-space perturbation energy excluded by truncation at degree $L$.

Because absolute perturbation energy does not indicate its magnitude relative to the geometric signal represented at the same spectral scale, we additionally calculated the perturbation-to-clean energy ratio
\begin{equation}
Q_{\ell}
=
\frac{
E_{\ell}^{\Delta}
}{
E_{\ell}^{\mathrm{clean}}
},
\qquad
E_{\ell}^{\mathrm{clean}}
=
\sum_{m=-\ell}^{\ell}
\left\|
\mathbf{c}^{\mathrm{clean}}_{\ell m}
\right\|_2^2.
\label{eq:perturbation_clean_ratio}
\end{equation}
Thus, $R_L$ quantifies the cumulative fraction of perturbation preserved by truncation at degree $L$, whereas $Q_{\ell}$ quantifies the perturbation burden relative to the clean geometric signal at an individual harmonic degree. In implementation, the denominator of $Q_{\ell}$ was bounded below by machine epsilon to prevent numerical division by zero. Values of $Q_{\ell}>1$ indicate that perturbation energy exceeds the clean-shape energy represented at the same spectral degree. Because $Q_{\ell}$ is normalized by degree-specific clean energy, it was interpreted jointly with $R_L$ rather than as a measure of absolute perturbation magnitude.

\subsubsection{Predictive robustness of curvature-derived features}
\label{sec:predictive_perturbation_robustness}

To assess whether resolution-dependent perturbation transmission affected downstream geometric analysis, we evaluated curvature-based prediction using $92$ descriptors consistently available for both clean and perturbed reconstructions at all six SH degrees across the $324$ lymph-node surfaces. These descriptors comprised mean curvature (26), Gaussian curvature (13), Shape Index and Curvedness (28), curvature integrals (9), and curvature regions (16), with identical feature definitions and numerical settings applied to the paired surfaces.

The patient-grouped modeling framework described in Section~\ref{sec:mixed_resolution_modeling} was retained. For each outer split, preprocessing, regularization selection, and model fitting were performed using only the clean training data. The fitted model was then applied without refitting to both the clean and perturbed versions of the held-out lymph nodes, thereby isolating perturbation-induced changes in prediction. The change in discriminative performance was defined as
\begin{equation}
\Delta \mathrm{AUC}_{L}
=
\mathrm{AUC}^{\mathrm{pert}}_{L}
-
\mathrm{AUC}^{\mathrm{clean}}_{L},
\label{eq:perturbation_delta_auc}
\end{equation}
where negative values indicate reduced discriminative performance after perturbation.

Prediction-level stability was additionally quantified using the mean absolute probability shift,
\begin{equation}
\Delta p_L
=
\frac{1}{N}
\sum_{i=1}^{N}
\left|
\hat{p}^{\mathrm{pert}}_{i,L}
-
\hat{p}^{\mathrm{clean}}_{i,L}
\right|,
\label{eq:mean_probability_shift}
\end{equation}
together with the Pearson correlation $r_L$ between paired clean and perturbed prediction probabilities. Larger $\Delta p_L$ and lower $r_L$ indicate reduced predictive stability.

Confidence intervals for $\Delta\mathrm{AUC}_{L}$ and pairwise comparisons of AUC degradation between SH degrees were obtained using paired patient-level bootstrap resampling, with all lymph nodes from the same patient resampled jointly, followed by Benjamini–Hochberg correction for multiple comparisons.

\subsection{Independent external and cross-organ validation} \label{sec:external_method}

\subsubsection{Multicenter label-free lymph node replication} \label{sec:external_lymphnode_method}

All 2,644 external-cohort surfaces underwent the complete preprocessing pipeline and shape analysis described previously (Sections~\ref{sec:data_geometric_processing} and~\ref{sec:shape_analysis}), as well as topology validation (Section~\ref{sec:topological_computational_quality}). Subsequently, a simple random sample of 1,000 topology-eligible lymph nodes was drawn without replacement for the computational tests described in Section~\ref{sec:topological_computational_quality}. The external replication analysis of degree-dependent feature effects was conducted separately using the complete cohort of 2,644 lymph nodes.

Because pathological labels were unavailable, the external cohort was used to test replication of degree-dependent geometric effects rather than clinical prediction performance.

The internal cohort comprised 1,769 lymph nodes from 214 unique patients. All lymph nodes from the same patient were assigned to the same cross-validation fold and were resampled jointly in patient-level bootstrap analyses. The analysis was restricted \textit{a priori} to the three geometric feature families available in both cohorts: convexity (6 features), mean curvature (26 features), and central asymmetry (10 features). SH-energy features were not included because the corresponding feature set was not available for matched external replication.

The degree set was fixed before examination of the external cohort. It was selected to cover the dominant family-specific degrees identified in the leakage-safe internal model-selection analysis: $L=5$ for convexity, $L=10$ for mean curvature, and $L=34$ for central asymmetry. In the internal analysis, degree selection was performed using inner-cross-validation AUC within each outer training fold; these values were the modal selected degrees for their respective feature families. To enable matched cross-cohort evaluation, all three feature families were recomputed at each of the common degrees, $L=5$, $L=10$, and $L=34$, in both cohorts. The external cohort was therefore used to assess replication of prespecified degree-dependent geometric signatures rather than to re-select an optimal degree.

Within each cohort, measurements obtained at different reconstruction degrees were treated as repeated observations from the same lymph node. For each feature with complete measurements at all three degrees, an omnibus Friedman test was used to assess degree dependence. Pairwise degree effects were estimated for the three prespecified contrasts: $L=5$ versus $L=10$, $L=10$ versus $L=34$, and $L=5$ versus $L=34$, using two-sided Wilcoxon signed-rank tests. For each contrast, the paired difference was defined as the value at the first degree minus the value at the second degree. Effect magnitude and direction were summarized using the rank-biserial correlation. Median paired differences were accompanied by 95\% percentile bootstrap confidence intervals based on 10,000 resamples. Benjamini--Hochberg false-discovery-rate correction was applied separately within each cohort across the feature-level omnibus tests and across the pairwise tests.

Each internal degree effect was matched to its external counterpart by feature family, feature identity, and degree contrast. Concordance between the internal and external rank-biserial effect sizes was quantified using Spearman's correlation coefficient, with 95\% percentile bootstrap confidence intervals obtained from 10,000 resamples of the matched effect pairs. Directional replication was summarized as the proportion of matched non-zero effects having the same sign in both cohorts; a two-sided exact binomial test against a probability of 0.5 was used as a supplementary assessment. Family-specific concordance analyses were considered primary, whereas the pooled analysis across all matched effects was considered secondary because the three families contributed unequal numbers of effects. This feature-level replication framework was consistent with established approaches for assessing replicability across multiple features while controlling multiplicity \citep{bogomolov2018replicability}.

To directly assess transferability of the internally identified family-specific degree preferences, we performed a prespecified directional composite analysis. For each feature $j$ and each alternative degree $L$, the expected direction was fixed from the internal cohort as
\begin{equation}
s_{f,j,L}
=
\operatorname{sign}
\left(
\widehat{\Delta}^{\mathrm{internal}}_{f,j,L_f^\ast-L}
\right),
\end{equation}
where $L_f^\ast$ denotes the internally preferred degree for feature family $f$. The external paired contrast was then direction-aligned and scaled using internal reference variability:
\begin{equation}
D_{i,f,j,L}
=
s_{f,j,L}
\frac{x_{i,f,j,L_f^\ast}-x_{i,f,j,L}}
{\operatorname{MAD}^{\mathrm{internal}}_{f,j}},
\end{equation}
where $x_{i,f,j,L}$ is the feature value for lymph node $i$, and the median absolute deviation was calculated exclusively from the internal cohort across the three common degrees. Thus, neither the direction, scaling, nor preferred degree was determined using external data.

For each lymph node and family, the aligned feature-level contrasts were aggregated using their median:
\begin{equation}
C_{i,f,L}
=
\operatorname{median}_{j\in f}
D_{i,f,j,L}.
\end{equation}
A positive value of $C_{i,f,L}$ indicates that the external node supports the internally specified preferred degree relative to the alternative degree. For each family--contrast pair, the lymph-node-level composites were evaluated directly. The composite median and its 95\% percentile confidence interval were estimated using 10,000 nonparametric bootstrap resamples of lymph nodes. The positive-node percentage was defined as the proportion of lymph nodes for which $C_{i,f,L}>0$. One-sided Wilcoxon signed-rank tests against zero were used to test whether the family-level composite was directionally positive. Holm correction was applied across the six prespecified family--contrast tests. This composite analysis was intended to test transferability of the internal degree preference, not equivalence of effect magnitude between cohorts.

\subsubsection{Cross-organ validation in LIDC-IDRI lung nodules}
\label{sec:lidc_crossorgan_method}

The LIDC-IDRI cohort was analyzed as an independent cross-organ experiment to determine whether the resolution-dependent relationship between SH representation and downstream discriminative performance extended beyond axillary lymph node morphology. SH-based three-dimensional shape analysis has previously been applied to lung-nodule malignancy discrimination \citep{elbaz2011lung}; here, the purpose of the LIDC-IDRI experiment was specifically to evaluate whether predictive performance remained dependent on SH reconstruction resolution in a distinct anatomical structure and classification task. For the supervised classification task, only nodules with definite diagnostic labels were considered. Label 1 nodules (benign/non-malignant) and label 2 nodules (primary malignant) were retained, whereas label 0 nodules with unknown diagnosis and label 3 metastatic lesions were excluded to maintain a clinically well-defined binary classification task. After further requiring usable CT data and corresponding segmentation contours suitable for three-dimensional surface reconstruction, 59 lung nodules from 56 unique patients were retained for analysis. Of these, 50 satisfied the strict topological criteria before geometric processing, while all 59 were retained for application of the complete preprocessing pipeline.

Following geometric processing and spherical parameterization, each lung-nodule surface was reconstructed using the prespecified SH degrees $L\in\{5,10,20,30,40,50\}$. Degree-specific representations were generated independently using the SH reconstruction framework described in Section~\ref{sec:spherical_harmonic_representation}.

For each SH degree, the same predefined feature-extraction procedure was applied to three geometric feature families: convexity, mean curvature, and central asymmetry. Degree-specific classifiers followed the same preprocessing and L2-regularized logistic-regression framework described in Section~\ref{sec:mixed_resolution_modeling}, with identical feature definitions and data partitions across all six degrees. Thus, SH reconstruction degree was the principal experimental factor differing among the compared representations.

Predictive performance was evaluated using five-fold outer cross-validation repeated three times, with all nodules from the same patient assigned to the same fold. Within each outer training set, the regularization parameter was selected by three-fold inner patient-grouped cross-validation using the same candidate grid described in Section~\ref{sec:mixed_resolution_modeling}. Identical outer partitions were retained across all six SH degrees.

Out-of-fold probabilities from the three outer-cross-validation repeats were averaged for each nodule. For each SH degree, discriminative performance was summarized using the area under the receiver-operating-characteristic curve (AUC). Ninety-five percent confidence intervals were estimated using 2,000 bootstrap resamples of the averaged out-of-fold predictions. Degree-specific AUC differences were assessed using paired bootstrap resampling with 2,000 iterations.

This cross-organ replication analysis evaluated whether the resolution-dependent relationship between SH representation and predictive performance generalized to a distinct anatomical structure and classification task.

\section{Results}

\subsection{Geometric processing establishes a valid computational surface domain}
\label{sec:results_processing}

Strict topology validity increased from 78.18\% (1383 /1769) before processing to 100.00\% (1769/1769) afterward, with all reported component criteria reaching 100\% (Table~\ref{tab:topology_validation}). In the $500$-surface computational-quality subset, processing increased fifth-percentile triangle quality and markedly reduced the 95th-percentile adjacent-face normal jump (Table~\ref{tab:computational_quality}). These results show that the processing workflow established a topology-valid and more regular computational surface domain for subsequent spherical and differential-geometric analysis.

\begin{table*}[!t]
\centering
\small
\caption{\textbf{Topological validity before and after geometric processing.}}
\label{tab:topology_validation}
\begin{tabularx}{\textwidth}{
>{\raggedright\arraybackslash}X
>{\centering\arraybackslash}c
>{\centering\arraybackslash}c
>{\centering\arraybackslash}c}
\hline
Criterion & Original & Processed & Change (pp) \\
\hline
Strict topology pass & 1383/1769 (78.18\%) & 1769/1769 (100.00\%) & +21.82 \\
Single-component surface & 1531/1769 (86.55\%) & 1769/1769 (100.00\%) & +13.45 \\
Genus-$0$ topology & 1537/1769 (86.89\%) & 1769/1769 (100.00\%) & +13.11 \\
Outward orientation & 1560/1769 (88.19\%) & 1769/1769 (100.00\%) & +11.81 \\
\hline
\end{tabularx}
\end{table*}

\begin{table*}[!t]
\centering
\small
\caption{\textbf{Downstream computational surface quality before and after geometric processing (internal cohort).} Values are medians across 500 paired surfaces.}
\label{tab:computational_quality}
\begin{tabularx}{\textwidth}{
>{\raggedright\arraybackslash}X
>{\centering\arraybackslash}c
>{\centering\arraybackslash}c
>{\centering\arraybackslash}X
>{\centering\arraybackslash}c}
\hline
Measure & Original & Processed & Median paired change (95\% CI) & FDR-adjusted $P$ \\
\hline
Triangle quality (5th percentile) & 0.7274 & 0.7787 & +0.0453 (0.0413--0.0506) & $2.29\times10^{-54}$ \\
Normal jump (95th percentile, degrees) & 38.7617 & 4.6759 & $-33.3190$ ($-33.9196$--$-32.7360$) & $5.06\times10^{-83}$ \\
\hline
\end{tabularx}
\end{table*}

\subsection{Multi-resolution spherical harmonic analysis}
\label{sec:results_multiresolution}

\subsubsection{Conventional shape descriptors maintain stable discrimination across SH representations}
\label{sec:conventional_shape_performance}

\begin{figure*}
\centering
\includegraphics[width=0.9\linewidth]{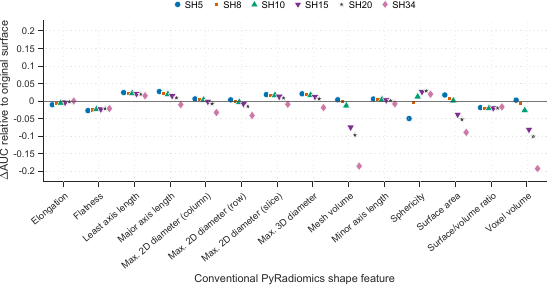}
\caption{\textbf{Changes in univariate discrimination of conventional PyRadiomics shape descriptors across SH representations.} For each descriptor and SH degree, $\Delta\mathrm{AUC}$ was calculated relative to the corresponding descriptor extracted from the original surface. Positive values indicate higher discrimination after SH reconstruction, whereas negative values indicate lower discrimination.}
\label{fig:py14_single_feature}
\end{figure*}

Changes in univariate \textit{Py14} discrimination were small across most SH resolutions (Fig.~\ref{fig:py14_single_feature}). Median descriptor-level $\Delta\mathrm{AUC}$ remained close to zero from degrees $5$ to $20$, with a modest reduction at degree $34$. Descriptor-specific responses were bidirectional, and Sphericity showed the largest degree dependence. Overall, conventional global-morphology discrimination was largely preserved after SH reconstruction.

\subsubsection{Degree-wise spectral discrimination is strongest at intermediate harmonic degrees}
\label{sec:degreewise_spectral_discrimination}

\begin{figure}
\centering
\includegraphics[width=\linewidth]{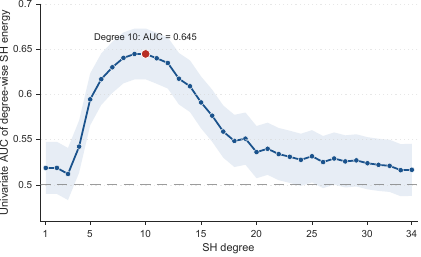}
\caption{\textbf{Degree-wise spectral discrimination across the SH spectrum.} Folded univariate AUC is shown for relative spectral energy at each harmonic degree, summarizing discrimination irrespective of the direction of association. Discrimination was strongest at intermediate degrees and declined toward chance level at the highest degrees.}
\label{fig:degreewise_spectral_discrimination}
\end{figure}

Degree-wise spectral analysis (Fig.~\ref{fig:degreewise_spectral_discrimination}) showed that relative spectral energy at intermediate harmonic degrees had the strongest univariate association with metastatic status. Discrimination was weak in the lowest harmonic components, increased progressively toward the intermediate-frequency range, and reached its strongest level around degrees $8$--$12$. The strongest individual degree was $\ell=10$, with a folded AUC of $\mathrm{AUC}^{*}=0.645$, while similar discrimination was observed at degrees $8$ and $9$. Beyond this intermediate range, degree-wise discrimination progressively declined, with folded AUC decreasing to 0.591 at degree $15$ and 0.536 at degree $20$ and approaching chance level in the highest harmonic components.

\subsubsection{Geometric fidelity increases monotonically, whereas predictive utility does not}
\label{sec:multiresolution_performance}

\begin{figure}
\centering
\includegraphics[width=\linewidth]{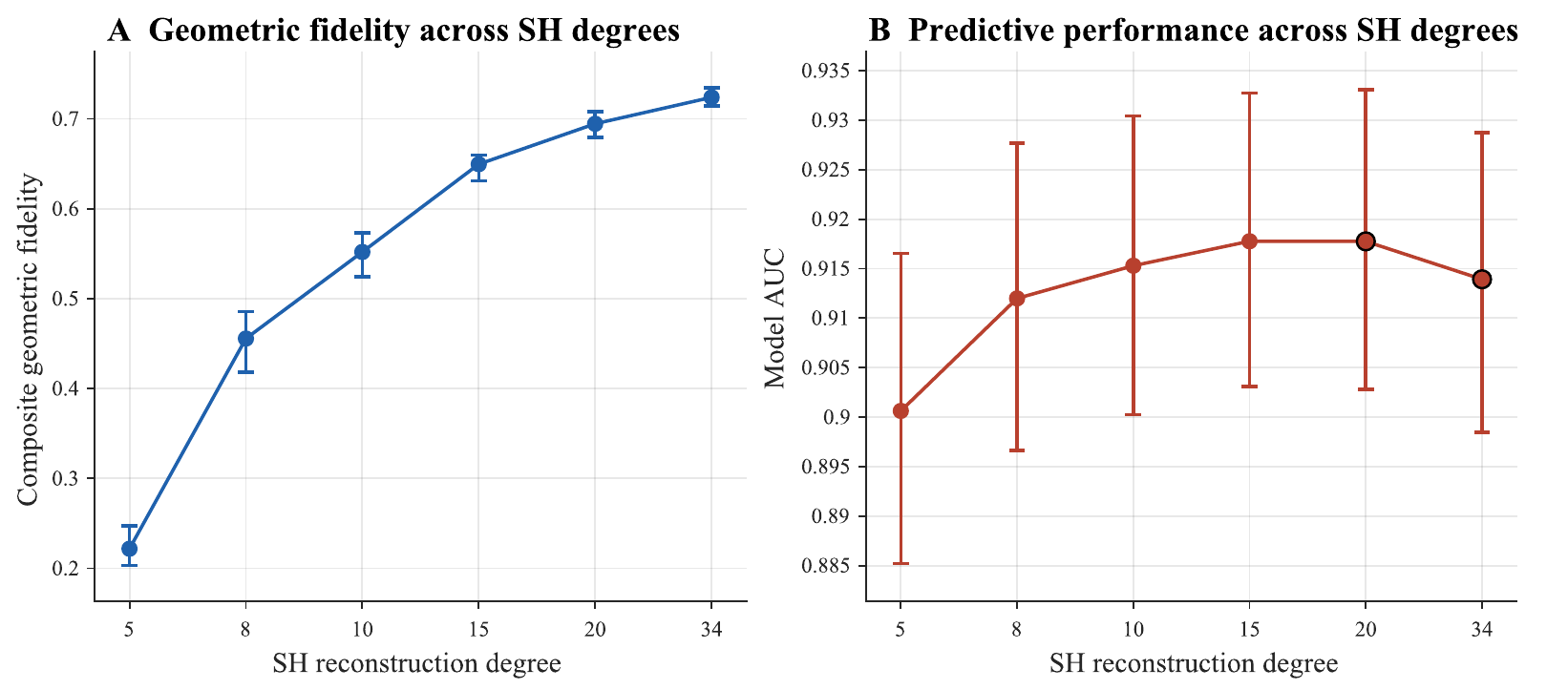}
\caption{\textbf{Divergence between geometric fidelity and predictive performance across SH degrees.} \textbf{(A)} Composite geometric fidelity increased monotonically with reconstruction degree. \textbf{(B)}~AUC of the fixed-resolution 366-feature models increased from low resolution to intermediate degrees but declined at degree $34$. Error bars denote 95\% confidence intervals.}
\label{fig:degree_fidelity_auc}
\end{figure}

Composite geometric fidelity increased monotonically with SH degree and was highest at $L=34$ (Fig.~\ref{fig:degree_fidelity_auc}A). Predictive performance followed a different pattern: fixed-resolution model AUC increased from low resolution to the intermediate degrees and subsequently decreased at $L=34$ (Fig.~\ref{fig:degree_fidelity_auc}B). Moreover, recovery of conventional shape measurements showed only a weak association with discriminative strength across feature--degree combinations (Spearman $\rho\approx0.25$). Numerical reconstruction fidelity and metastasis discrimination therefore followed distinct resolution-dependent behaviors.

\subsubsection{Feature-dependent SH resolution preferences}
\label{sec:results_family_resolution}

\begin{figure}
\centering
\includegraphics[width=\linewidth]{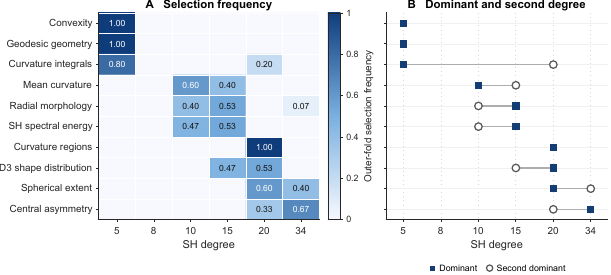}
\caption{\textbf{Feature-family-specific SH degree selection.}
\textbf{(A)}~Selection frequencies are shown across the locked outer cross-validation folds for ten representative feature families; degree selection was performed for all 20 predefined feature families.
\textbf{(B)}~Distinct descriptor families favored different SH reconstruction degrees, demonstrating that a single common resolution was not uniformly optimal across geometric measurements.}
\label{fig:family_degree_selection}
\end{figure}

The preferred SH degree varied substantially across the predefined geometric feature families (Fig.~\ref{fig:family_degree_selection}). Many curvature-related, spectral-energy, and shape-distribution descriptors favored intermediate resolutions, whereas several global and intrinsic descriptors favored lower degrees and selected families favored the highest resolution. These heterogeneous selection patterns demonstrate that the reconstruction scale most informative for prediction depends on the geometric quantity being measured.

The family-specific mixed-resolution model achieved the highest overall predictive performance among the evaluated representations (Table~\ref{tab:multiresolution_model_performance}). Its performance was comparable to the fixed-resolution models at degrees $8$, $10$, $15$, $20$, and $34$ after multiple-comparison correction, but significantly exceeded the low-resolution degree-$5$ model. Family-specific selection therefore primarily avoided clearly mismatched representation scales rather than substantially improving performance beyond the best intermediate fixed resolution.

\begin{table}[!t]
\centering
\small
\caption{\textbf{Predictive performance of fixed- and mixed-resolution SH representations.}}
\label{tab:multiresolution_model_performance}
\resizebox{\linewidth}{!}{$
\begin{tabular}{lcC{2cm}c}
\hline
Representation & AUC (95\% CI) & Mixed minus model $\Delta$AUC & FDR-adjusted $P$ \\
\hline
SH degree 5 & 0.9006 (0.8852--0.9165) & 0.0177 & 0.006 \\
SH degree 8 & 0.9120 (0.8966--0.9277) & 0.0064 & 0.384 \\
SH degree 10 & 0.9153 (0.9002--0.9304) & 0.0031 & 0.588 \\
SH degree 15 & 0.9178 (0.9031--0.9327) & 0.0006 & 0.817 \\
SH degree 20 & 0.9178 (0.9028--0.9331) & 0.0006 & 0.817 \\
SH degree 34 & 0.9139 (0.8985--0.9288) & 0.0044 & 0.504 \\
Mixed resolution & \textbf{0.9183 (0.9037--0.9336)} & -- & -- \\
\hline
\end{tabular}
$}
\end{table}

Compared with the original-surface \textit{Py14} baseline, the mixed-resolution model improved the AUC from 0.884 (95\% CI, 0.868--0.900) to 0.918 (95\% CI, 0.904--0.934), corresponding to an absolute improvement of 0.0344 (95\% paired bootstrap CI, 0.0212--0.0468; $P=0.001$). This improvement indicates that the extended geometric descriptors provide discriminative information complementary to conventional global morphology.

\subsubsection{High-resolution SH achieves compact geometric representation with high fidelity}
\label{sec:results_high_resolution_efficiency}

Despite its lower predictive utility relative to intermediate resolutions, degree $34$ provided the closest approximation to the processed reference geometry. In the $500$-lymph-node fidelity subset, the median normalized symmetric mean nearest-neighbor distance and HD95 were 0.00887 and 0.01744, respectively, while relative surface-area and volume errors were 0.120\% and 0.006\%. Agreement of the 14 \textit{Py14} descriptors with the reference surfaces was also high (median Pearson, Spearman, and Lin's concordance correlations: 0.9995, 0.9999, and 0.9995). A degree-$34$ representation used $1,225$ three-dimensional coefficient vectors compared with a median of $38,492$ mesh vertices, corresponding to a 31.42-fold reduction in representation elements. Thus, high-resolution SH remained useful when faithful and structured geometric representation, rather than maximal discrimination, was the primary objective.

\subsection{Resolution-dependent perturbation robustness}

\begin{figure*}[!t]
\centering
\includegraphics[width=0.65\textwidth]{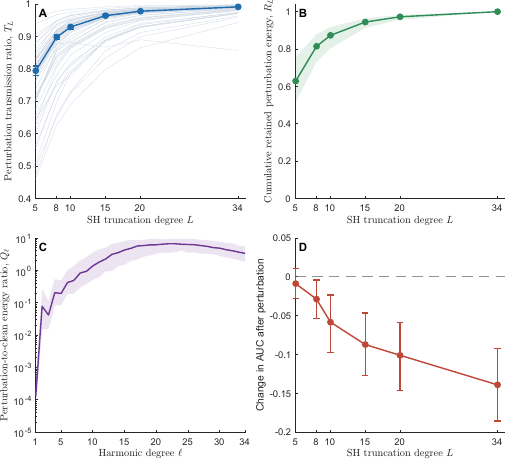}
\caption{\textbf{Resolution-dependent perturbation robustness across SH degrees.}
\textbf{(A)}~Geometric perturbation transmission ratio $T_L$.
\textbf{(B)}~Cumulative retained perturbation energy $R_L$.
\textbf{(C)}~Perturbation-to-clean spectral energy ratio $Q_{\ell}$.
\textbf{(D)}~Perturbation-induced change in AUC. Complementary predictive-stability measures included the mean absolute probability shift
$\Delta p_L$ and clean--perturbed prediction correlation.}
\label{fig:spectral_perturbation_robustness}
\end{figure*}

Geometric perturbation transmission increased with SH degree (Fig.~\ref{fig:spectral_perturbation_robustness}A), with the median transmission ratio rising from $T_5=0.794$ to $T_{34}=0.992$. Thus, progressively higher-resolution representations retained a larger fraction of the imposed surface displacement.

The harmonic-domain analysis showed the same resolution dependence. Cumulative retained perturbation energy increased with truncation degree (Fig.~\ref{fig:spectral_perturbation_robustness}B), while the perturbation-to-clean energy ratio exceeded unity from approximately the intermediate harmonic range and remained elevated at high degrees (Fig.~\ref{fig:spectral_perturbation_robustness}C). These results indicate that increasing SH bandwidth retains progressively more perturbation energy and increases its relative contribution at fine spectral scales.

The same pattern propagated to curvature-based prediction (Fig.~\ref{fig:spectral_perturbation_robustness}D). Perturbation-induced AUC degradation was significantly greater than that at $L=5$ for degrees $10$, $15$, $20$, and $34$ after FDR correction. Across the 324 lymph nodes, higher SH degrees produced larger mean absolute probability shifts and lower clean--perturbed prediction correlations.

\subsection{Metastasis-associated geometric phenotypes}

Beyond aggregate model performance, representative descriptors from the interpretable geometric families showed substantial separation between metastatic and non-metastatic lymph nodes. The strongest examples included lower-tail curvedness and absolute mean-curvature descriptors, Gaussian-curvature statistics, radial morphology, and convexity-related measurements, with univariate AUCs ranging from 0.824 to 0.898. The corresponding Cliff's $\delta$ values indicated large effect sizes across both global and local geometric properties.

These descriptors characterize different manifestations of the nodal surface phenotype. Radial and convexity-related measurements quantify departures from a regular overall contour, whereas curvedness and curvature statistics capture regional variations in surface bending and contour irregularity. Such geometric changes may be consistent with structural remodeling accompanying metastatic involvement, including asymmetric cortical thickening, hilar effacement, or heterogeneous tumor infiltration, which can alter the external contour of the lymph node. Their joint association with metastatic status therefore suggests that relevant morphological information extends beyond global enlargement or compactness to include spatially heterogeneous surface remodeling. Together, these measurements provide an interpretable quantitative description of metastasis-associated nodal morphology complementary to conventional global radiomic shape features.

\subsection{Independent external and cross-organ validation}
\label{sec:external_result}

\subsubsection{Multicenter label-free lymph node replication}
\label{sec:external_lymphnode_result}

The multicenter external cohort comprised 2,644 lymph nodes without node-level pathological metastasis labels and was used to assess label-free replication of the geometric-processing and family-specific degree effects.

Geometric processing established complete strict topology validity in both independent datasets (Table~\ref{topology_result}). All 2,644 multicenter lymph node surfaces and all 59 LIDC-IDRI lung-nodule surfaces satisfied the predefined topological requirements after processing, with a larger pre-processing deficit observed in the LIDC-IDRI cohort.

\begin{table*}[htbp]
\centering
\caption{\textbf{Topological validity before and after geometric processing in the independent external and cross-organ cohorts.}}
\label{topology_result}
\setlength{\tabcolsep}{8pt}
\resizebox{\linewidth}{!}{$
\begin{tabular}{llccc}
\toprule
Dataset & Criterion & Original & Processed & Change (percentage points) \\
\midrule
External ($n=2644$)
& Strict topology pass
& 2607/2644 (98.60\%)
& 2644/2644 (100.00\%)
& +1.40 \\

& Single-component surface
& 2623/2644 (99.21\%)
& 2644/2644 (100.00\%)
& +0.79 \\

& Genus-0 topology
& 2628/2644 (99.40\%)
& 2644/2644 (100.00\%)
& +0.61 \\

& Outward orientation
& 2644/2644 (100.00\%)
& 2644/2644 (100.00\%)
& +0.00 \\

\midrule
LIDC-IDRI ($n=59$)
& Strict topology pass
& 50/59 (84.75\%)
& 59/59 (100.00\%)
& +15.25 \\

& Single-component surface
& 58/59 (98.31\%)
& 59/59 (100.00\%)
& +1.69 \\

& Genus-0 topology
& 50/59 (84.75\%)
& 59/59 (100.00\%)
& +15.25 \\

& Outward orientation
& 58/59 (98.31\%)
& 59/59 (100.00\%)
& +1.69 \\

\bottomrule
\end{tabular}
$}
\end{table*}

Computational-quality measures also improved substantially after geometric processing in both independent datasets (Table~\ref{computation_result}). Processing increased lower-tail triangle quality and markedly reduced extreme adjacent-face normal variation in both the multicenter lymph node and LIDC-IDRI cohorts, indicating improved mesh regularity across distinct anatomical datasets.

\begin{table*}[htbp]
\centering
\caption{\textbf{Paired computational-quality results before and after geometric processing in the independent external and cross-organ cohorts.} Differences are processed minus original values; confidence intervals were obtained by paired bootstrap resampling.}
\label{computation_result}
\setlength{\tabcolsep}{6pt}
\resizebox{\linewidth}{!}{$
\begin{tabular}{llrrrr}
\toprule
Dataset & Metric & Original median & Processed median &
Median difference (95\% CI) & FDR-adjusted $q$ \\
\midrule
External ($n=1000$)
& Triangle quality (5th percentile)
& 0.678 & 0.769
& +0.088 (0.082 to 0.097)
& $<10^{-123}$ \\

& Normal jump (95th percentile, degrees)
& 32.74 & 3.38
& -29.45 (-29.89 to -29.01)
& $<10^{-164}$ \\

\midrule
LIDC-IDRI ($n=59$)
& Triangle quality (5th percentile)
& 0.536 & 0.753
& +0.178 (0.156 to 0.235)
& $3.59\times10^{-11}$ \\

& Normal jump (95th percentile, degrees)
& 68.20 & 4.92
& -63.16 (-64.67 to -60.38)
& $3.59\times10^{-11}$ \\

\bottomrule
\end{tabular}
$}
\end{table*}

The label-free replication analysis was restricted \textit{a priori} to the three feature families available in both lymph node cohorts: convexity (6 features), mean curvature (26 features), and central asymmetry (10 features). Their internally specified preferred degrees were $L=5$, $L=10$, and $L=34$, respectively. All three families were therefore recomputed at the common degrees $L=5$, $L=10$, and $L=34$ in both cohorts.

The external cohort reproduced the family-specific degree-effect signatures observed internally (Table~\ref{compare_external}). Feature-level effect sizes showed strong internal--external concordance across all three families, with a pooled Spearman correlation of $\rho=0.964$ and directional agreement for 121 of 126 matched effects. Both the direction and relative ordering of degree-dependent feature effects were therefore largely preserved across cohorts.

\begin{table}[htbp]
\centering
\caption{\textbf{Concordance of feature-level degree effects between the internal and independent external cohorts.} Matched effects comprise all feature-by-degree-contrast pairs ($L5$ vs $L10$, $L10$ vs $L34$, and $L5$ vs $L34$). Spearman's $\rho$ quantifies concordance between internal and external paired rank-biserial effect sizes; direction agreement is the number of matched effects with the same sign.}
\label{compare_external}
\resizebox{\linewidth}{!}{$
\begin{tabular}{lC{2cm}rl}
\toprule
Feature family & Matched effects & Spearman $\rho$ (95\% CI) & Direction agreement \\
\midrule
Convexity & 18 & 0.872 (0.569--0.987) & 18/18 \\
Central asymmetry & 30 & 0.942 (0.830--0.986) & 30/30 \\
Mean curvature & 78 & 0.979 (0.956--0.988) & 73/78 \\
\midrule
Pooled & 126 & 0.964 (0.941--0.977) & 121/126 \\
\bottomrule
\end{tabular}
$}
\end{table}

The prespecified direction-aligned family composites further supported transfer of the internally identified degree preferences (Table~\ref{family_degree_transfer}). All six family--contrast composites were positive, their 95\% bootstrap confidence intervals excluded zero, and all comparisons remained significant after Holm correction. The strength of replication nevertheless differed across feature families, with the weakest separation observed for central asymmetry between $L=34$ and $L=10$.

\begin{table*}[htbp]
\centering
\small
\caption{\textbf{External replication of internally specified family-specific degree preferences.} Composite contrasts were direction-aligned using feature-specific directions defined in the internal cohort and scaled using internal median absolute deviations. Positive values support the internally preferred degree relative to the alternative degree. Confidence intervals were obtained using 10,000 nonparametric bootstrap resamples of lymph nodes. Positive lymph nodes are those with a direction-aligned composite greater than zero. One-sided $P$ values were adjusted across the six prespecified family--contrast tests using the Holm procedure.}
\label{family_degree_transfer}
\resizebox{\linewidth}{!}{
\begin{tabular}{lccrrrrrr}
\toprule
Feature family &
Preferred $L$ &
Alternative $L$ &
Features &
Composite median &
95\% CI &
Positive lymph nodes (\%) &
One-sided $P$ &
Holm-adjusted $P$ \\
\midrule
Convexity &
5 & 10 & 6 &
0.116 & 0.110--0.122 & 98.6 &
$<0.001$ & $<0.001$ \\

Convexity &
5 & 34 & 6 &
0.223 & 0.213--0.232 & 99.2 &
$<0.001$ & $<0.001$ \\

Mean curvature &
10 & 5 & 26 &
0.216 & 0.210--0.220 & 98.8 &
$<0.001$ & $<0.001$ \\

Mean curvature &
10 & 34 & 26 &
0.970 & 0.943--1.004 & 99.7 &
$<0.001$ & $<0.001$ \\

Central asymmetry &
34 & 5 & 10 &
0.166 & 0.154--0.174 & 79.4 &
$<0.001$ & $<0.001$ \\

Central asymmetry &
34 & 10 & 10 &
0.024 & 0.021--0.028 & 65.4 &
$<0.001$ & $<0.001$ \\

\bottomrule
\end{tabular}
}
\end{table*}

Together, these results support the transportability of the geometric-processing framework and the family-specific multi-resolution effects to the independent multicenter lymph node cohort.

\subsubsection{Family-specific favored SH degrees in LIDC-IDRI lung nodules}
\label{sec:lidc_crossorgan_result}

The labeled LIDC-IDRI experiment showed feature-family-specific dependence of predictive performance on SH degree (Table~\ref{tab:lidc_family_favoured_degree}). Convexity achieved its highest observed AUC at $L=5$, whereas mean curvature and central asymmetry favored $L=40$ and $L=30$, respectively. Despite the distinct anatomy and classification endpoint, predictive utility therefore remained resolution-dependent, and the favored degree varied across feature families.

\begin{table*}[htbp]
\centering
\small
\caption{\textbf{Observed family-specific SH-degree performance in the
LIDC-IDRI cross-organ experiment.} Values are shown as AUC with 95\%
confidence intervals. Paired $\Delta$AUC values denote the differences between the highest and second-highest observed AUCs and were calculated using the same out-of-fold predictions.}
\label{tab:lidc_family_favoured_degree}
\begin{tabular}{lccc}
\toprule
Feature family &
Highest observed degree &
AUC (95\% CI) &
Paired $\Delta$AUC (95\% CI) \\
\midrule
Convexity &
${L=5}$ &
${0.853}$ (0.731--0.945) &
0.013 (-0.073--0.093) \\

Mean curvature &
${L=40}$ &
${0.883}$ (0.788--0.964) &
0.063 (-0.012--0.142) \\

Central asymmetry &
${L=30}$ &
${0.745}$ (0.606--0.879) &
0.002 (-0.034--0.042) \\
\bottomrule
\end{tabular}
\end{table*}

\section{Discussion}

This study shows that SH degree should be interpreted as a task-dependent geometric scale rather than a resolution parameter to be maximized uniformly. Across geometric fidelity, metastasis discrimination, perturbation robustness, and cross-organ prediction, different analytical objectives and anatomical settings favored different spectral resolutions.

\subsection{Reliable surface representation enables quantitative geometric analysis}

The increase in strict topology validity to 100\%, together with the marked improvement in triangle regularity and adjacent-face normal consistency, demonstrates that preprocessing is not merely a cosmetic smoothing step but establishes the computational domain required for subsequent spherical and differential-geometric analysis. This distinction is important because a nominal genus-$0$ classification alone does not guarantee the connectedness, manifoldness, orientation consistency, or local numerical regularity required by geometric operators \citep{brechbuehler1995parametrization,choi2015flash,meyer2003discrete}. The reproduction of these improvements in the independent multicenter cohort further indicates that the processing procedure was not specific to the internal data source.

\subsection{SH resolution defines a task-dependent geometric scale}

The divergence between reconstruction fidelity and discrimination is central to the interpretation of SH resolution. Increasing degree systematically improved geometric recovery, as expected from the hierarchical structure of the SH basis \citep{kazhdan2003}, but the strongest metastatic-status discrimination occurred at intermediate resolutions. High-frequency geometric content therefore cannot be assumed to be uniformly task-informative. In this setting, spectral truncation is better interpreted as scale selection than as simple information loss: low degrees may omit disease-associated regional morphology, whereas high degrees may increasingly retain fine contour variations arising from both anatomy and image- or segmentation-related variability. Intermediate resolutions may consequently provide a more favorable balance between morphological information and fine-scale uncertainty.

The preferred scale also depended on the geometric quantity being measured. Global morphology, curvature, intrinsic geometry, spatial distributions, and symmetry characterize different aspects of surface organization and therefore need not respond similarly to spectral bandwidth. Accordingly, the family-specific selection results support incorporating representation scale into feature construction rather than imposing a single global reconstruction degree. The mixed-resolution model was comparable to the best intermediate fixed-degree models, suggesting that family-specific selection primarily avoids mismatched representation scales across heterogeneous geometric measurements.

Previous medical applications have demonstrated the utility of SH-derived shape information for lung-nodule and cardiac classification \citep{elbaz2011lung,valizadeh2021parametric}. The LIDC-IDRI experiment extended this observation to a distinct anatomical structure and classification task. Despite substantial differences between lung nodules and axillary lymph nodes, predictive performance remained dependent on SH reconstruction degree. These results further support a task-dependent interpretation of spectral resolution, with the preferred bandwidth potentially varying with anatomy, feature definition, and downstream objective.

\subsection{Resolution-dependent perturbation sensitivity and the fidelity--robustness trade-off}

The perturbation experiments provide a complementary explanation for the resolution-dependent behavior of SH representations. Spectral truncation attenuated a larger fraction of the imposed geometric variation at lower degrees, whereas higher degrees preserved progressively more of the perturbation in both the reconstructed geometry and harmonic coefficients. The same trend propagated to curvature-derived predictions, whose stability decreased as resolution increased. Because local differential quantities are particularly responsive to fine-scale surface variation, the additional bandwidth that improves high-resolution geometric fidelity can simultaneously increase sensitivity to perturbations.

Higher geometric fidelity and lower perturbation attenuation are two consequences of increased representational bandwidth. Preserving finer geometry necessarily reduces spectral suppression of small-scale variation. Consequently, the appropriate SH degree depends on both the level of geometric detail to be retained and the stability requirements of the downstream measurement.

\subsection{Limitations and future directions}
The present framework has two main limitations. First, it relies on the fidelity of CT-derived surface reconstruction. Finite spatial resolution, anisotropic sampling, partial-volume effects, and segmentation variability may introduce geometric uncertainty that cannot be fully separated from true anatomical variation. This issue is particularly relevant for curvature-based descriptors, which are sensitive to local geometric perturbations and discretization effects \citep{meyer2003discrete,gatzke2006estimating}. Although the proposed preprocessing improves mesh validity and numerical stability, it cannot recover anatomical information absent from the original imaging data.

Second, SH degree selection provides only an indirect mechanism for controlling the trade-off between geometric fidelity and robustness. Lower-degree representations suppress high-frequency variations but may remove informative geometry, whereas multi-resolution representations increase computational cost. Future work should investigate differential-response-based representations that directly characterize and regulate the sensitivity of geometric measurements to surface perturbations.

\section{Conclusion}

We developed an integrated framework for topology-aware processing and multi-resolution geometric analysis of CT-derived axillary lymph node surfaces. The proposed processing pipeline established a connected, closed, consistently oriented genus-$0$ computational domain for spherical analysis while improving local mesh quality and limiting unnecessary modification of the underlying geometry. This provides a reproducible geometric basis for applying differential-geometric and spectral descriptors to clinically derived three-dimensional surfaces.

The principal finding is that SH resolution should be regarded as a task-dependent geometric scale rather than as a reconstruction parameter that should be maximized uniformly. Increasing SH degree progressively improved reconstruction fidelity, but metastasis-associated discrimination was strongest at intermediate spectral scales, and different geometric feature families favored different resolutions. The family-specific mixed-resolution representation accommodated these heterogeneous scale preferences and captured discriminative information beyond conventional global shape descriptors, while high-resolution SH remained valuable for compact and faithful representation of detailed surface geometry. Thus, the resolution that best preserves the observed surface need not be the resolution that is most informative for a downstream prediction task.

Controlled perturbation analysis further showed that SH resolution also governs the transmission of fine-scale geometric variation: higher-degree representations preserved a larger fraction of the imposed perturbation, whereas spectral truncation provided greater attenuation and correspondingly greater stability of curvature-based predictions. Together, these findings distinguish geometric fidelity, discriminative utility, and perturbation robustness as complementary but non-equivalent objectives of quantitative shape representation. The proposed framework therefore provides a principled basis for selecting geometric resolution according to the downstream analytical objective and offers a general strategy for multi-scale quantitative characterization of lymph node morphology and other clinically derived three-dimensional anatomical surfaces.

\printcredits

\bibliographystyle{cas-model2-names}

\bibliography{reference}

\end{document}